\documentclass[11pt]{article}
\usepackage{graphics}
\usepackage{epsfig}

\def\be{\begin{equation}}  
\def\bea{\begin{eqnarray}}  
\def\ee{\end{equation}}     
\def\eea{\end{eqnarray}}     
\def\gsim{\mathrel{\raise.3ex\hbox{$>$\kern-.75em\lower1ex\hbox{$\sim$}}}}
\def\lsim{\mathrel{\raise.3ex\hbox{$<$\kern-.75em\lower1ex\hbox{$\sim$}}}}

\begin{document}

\vskip .7cm

\vspace{30cm}

\begin{center}

{\Large \bf  Cosmology with mirror dark matter II: \\
\vspace{0.3cm}
Cosmic Microwave Background and Large Scale Structure }

\vskip .7cm

{\large Paolo Ciarcelluti}

\vskip .7cm

{\it Dipartimento di Fisica, Universit\`a di L'Aquila, 67010 Coppito  
AQ, and  \\  INFN, Laboratori Nazionali del Gran Sasso, 67010 Assergi AQ, 
Italy \\ 
E-mail: {\tt ciarcelluti@lngs.infn.it}
\vspace{8pt}}

\end{center}

\vskip .3cm

\begin{abstract}

This is the second paper of a series devoted to the study of the 
cosmological implications of the existence of mirror dark matter. 
The parallel hidden mirror world has the same microphysics 
as the observable one and couples the latter only gravitationally. 
The primordial nucleosynthesis bounds demand that the mirror sector 
should have a smaller temperature $ T' $ than the ordinary one $ T $, 
and by this reason its evolution can be substantially deviated from the 
standard cosmology. 
In this paper we took scalar adiabatic perturbations as the input in a flat 
Universe, and computed the power spectra for ordinary and mirror 
CMB and LSS, changing the cosmological parameters, and always 
comparing with the CDM case. 
We found differences in both the CMB and LSS power spectra, and we 
demonstrated that the LSS spectrum is particularly sensitive 
to the mirror parameters, due to the presence of both the oscillatory 
features of mirror baryons and the collisional mirror Silk damping. 
For $ x < 0.3 $ the mirror baryon-photon decoupling happens before the 
matter-radiation equality, so that CMB and LSS power spectra in 
linear regime are equivalent for mirror and CDM cases.
For higher $x$-values the LSS spectra strongly depend on the amount of 
mirror baryons. 
Finally, qualitatively comparing with the present observational limits on the 
CMB and LSS spectra, we show that for $ x < 0.3 $ the entire dark matter 
could be made of mirror baryons, while in
the case $ x \gsim 0.3 $ the pattern of the LSS power spectrum excludes the 
possibility of dark matter consisting entirely of mirror baryons, but they 
could present as admixture (up to $ \sim 50\% $) to the conventional CDM. 

\end{abstract}

\vspace{0.3cm}


\section{Introduction}

For a more general introduction we send the reader to the first paper of 
this series \cite{paper1} (hereafter referred to as Paper I), where we studied 
the linear evolution of adiabatic scalar primordial perturbations.
Here we recall the basics of the mirror matter theory and the main results 
of previous Paper I.

Mirror matter is an ideal stable dark matter candidate.
The basic concept is to have a hidden mirror (M) sector of the Universe 
which has exactly the same particles and interactions of our observable (O) 
sector.\footnote{From now on all fields and 
quantities of the mirror sector will have an apex to distinguish them from the 
ones belonging to the observable or ordinary world.}
The theory asserts that a discrete symmetry $P(G\leftrightarrow G')$ 
interchanging corresponding fields of $G$ and $G'$, so-called mirror 
parity, guarantees that both particle sectors are described by the same 
Lagrangians\footnote{In the brane world picture, the M sector can be the 
same O world realized on a parallel brane. },
with all coupling constants (gauge, Yukawa, Higgs) having the same 
pattern, so that their microphysics is the same\footnote{ There is the 
possibility that the mirror parity could be spontaneously broken and the 
weak interaction scales $\langle \phi \rangle =v$ and 
$\langle \phi' \rangle =v'$ could be different, leading to somewhat different 
particle physics in the mirror sector \cite{broken}. 
In this paper we consider only the case $ v = v' $, where the physics is the 
same in both sectors. } \cite{mirror,blinkhlo}. 
Thus, mirror particles are stable exactly as their ordinary counterparts. 
In addition, two sectors communicate with each other essentially through 
gravity. 
Indeed, they could communicate perhaps also via some other 
messengers\footnote{
The possible non-gravitational interactions include photon--mirror-photon 
kinetic mixing, neutrino--mirror-neutrino mass mixing, 
and Higgs-boson--mirror-Higgs-boson mixing. }
\cite{mixing}, but they are controlled by free parameters, that we can 
choose small enough to neglect these interactions.

Many people were interested in mirror world over last years, 
in particular being motivated by the problems of 
neutrino physics \cite{neutrino}, 
gravitational microlensing \cite{Macho}, 
gamma ray bursts \cite{mir_GRB}, 
flavour and CP violation \cite{assione}
large scale structure of the Universe \cite{ignavol-lss,paolo}, 
galaxy formation \cite{mir_halo},
orthopositronium lifetime \cite{ortho,bader}, 
dark matter detection experiments \cite{mir_dama}, 
anomalous events within the solar system 
\cite{mir_meteor,detect_mir_frag,pioneer}, 
extrasolar planets \cite{mir_planet}.

Then, if the mirror hypothesis is verified, the Universe contains, besides 
the particles of the ordinary sector, also their partners in the mirror sector.
However, the fact that there are the same particles and interactions in both 
sectors does not mean that also their cosmological densities are the same. 
In fact, two sectors could have different initial conditions, and this is exactly 
what is required in order to avoid a conflict with the Big Bang 
nucleosynthesis (BBN) bounds on  the effective number of extra light 
neutrino species. 
In particular, the BBN bounds imply that the M sector has a temperature 
lower than the O one, with a ratio $ x = T' / T < 0.64 $ \cite{bcv}. 
We remark that the sensitivity of this limit to the factor $ x $ is very low, 
so that also values $ x < 0.7 $ could be considered.
In order to obtain this situation, the following conditions must be satisfied:

A. After the Big Bang the two systems are born with different temperatures, 
namely the post-inflationary reheating temperature in the M sector is lower 
than in the visible one, $T'_R < T_R$. 
This can be naturally achieved in certain models \cite{inflation}.

B. The two systems interact very weakly, so that they do not come into 
thermal equilibrium with each other after reheating.
This condition is automatically fulfilled  if the two worlds communicate only 
via gravity.
If there are some other effective couplings between the O and M particles, 
they have to be properly suppressed. 

C. Both systems expand adiabatically, without significant entropy production.
If the two sectors have different reheating temperatures, during the 
expansion of the Universe they evolve independently, their temperatures 
remain different at later stages, $T' < T$, and the presence of the M sector 
would not affect primordial nucleosynthesis in the ordinary world.  

As shown in previous papers \cite{paper1, paolo, bcv}, due to the 
temperature difference between the two sectors, the key epochs for 
structure formation take place at different redshifts, and in particular they 
happens in the M sector before than in the O one. 
We define the two free parameters describing the mirror world as:
\be \label{mir-par}
x = {T' \over T} \lsim 0.64
\;\;\;\;\;\;\;\;\;\;\;\;\;\;\;\;\;\; 
\beta = {\Omega'_b \over \Omega_b} \gsim 1 \; ,
\ee
where the first limit comes from the aforementioned BBN bounds, while the 
second one from the hypothesis that mirror matter could constitute an 
important fraction of dark matter in the Universe. 
This latter assumption is supported by various baryogenesis scenarios with 
a hidden sector \cite{baryo-lepto}, all obtaining a mirror baryonic density at 
least equal to the ordinary one. 

Then, in a so-called Mirror Universe (a misleading expression indicating a 
Universe made of two sectors, ordinary and mirror) the radiation and matter 
components are expressed in general by
\be
\Omega_{r} = (\Omega_{r})_{\rm O} + (\Omega_{r})_{\rm M} = 
  (\Omega_{r})_{\rm O} (1 + x^4) = 4.2 \times 10^{-5}\,h^{-2}\,(1+x^4) \; ,
\ee
where the additional term $ x^4 $ takes into account the temperature 
difference between two sectors and is very low in view of the bound 
(\ref{mir-par}), 
and
\be
\Omega_m = \Omega_b + \Omega'_b + \Omega_{CDM}
                   = \Omega_b (1 + \beta) + \Omega_{CDM} \; ,
\ee
where $ \Omega'_b $ is the contribution of mirror baryonic dark matter 
(MBDM) to the density.

The relevant epochs for structure formation are matter-radiation equality 
(MRE), matter-radiation decoupling (MRD) and photon-baryon equipartition, 
respectively represented by the redshifts: 
\be \label{z-eq} 
1+z_{\rm eq}= {{\Omega_m} \over {\Omega_r}} \approx 
 2.4\cdot 10^4 {{\Omega_{m}h^2} \over {1+x^4}} \;,~~
\ee
\be \label{z'_dec}
1+z'_{\rm dec} \simeq x^{-1} (1+z_{\rm dec}) 
\simeq 1.1\cdot 10^3 x^{-1} \;, ~~
\ee
\be \label{shiftzbg} 
1+z_{\rm b\gamma}' 
  = { \Omega_b' \over \Omega_{\gamma}' } 
  \simeq { \Omega_b \, \beta \over \Omega_\gamma \, x^4} 
  = (1+z_{\rm b\gamma}) { \beta \over x^4 } > 1+z_{\rm b\gamma} \;.
\ee
Eq.~(\ref{z-eq}) means that the MRE in a Mirror Universe occurs always 
later than in an ordinary one, while from eqs.~(\ref{z'_dec}) and 
(\ref{shiftzbg}) we know that decoupling and equipartition in the M sector 
happen before than in the O one.
If we plot the redshifts of MRE and mirror MRD as functions of $ x $, 
as in fig.~1 of ref.~\cite{paper1}, we obtain their intersection for a 
value $ x_{\rm eq} $ expressed by
\be
x_{\rm eq} \approx 0.046(\Omega_m h^2)^{-1} \; .
\ee
This means that for values $ x < x_{\rm eq} $ the mirror decoupling occurs 
in the radiation dominated period, with consequences on the structure 
formation process (see Paper I \cite{paper1} and 
refs.~\cite{ignavol-lss,paolo,bcv}).
Using the value $\Omega_m h^2 = 0.135$, appeared in a recent WMAP fit 
\cite{wmap-par}, we obtain the typical value $ x_{\rm eq} \approx 0.34 $. 

This quantity has a key role in structure formation with mirror dark matter.
In fact we know from the results of Paper I \cite{paper1} and 
refs.~\cite{ignavol-lss,paolo}) that for values 
$ x < x_{\rm eq} $ the evolution of primordial perturbations in the linear 
regime is practically identical to the standard CDM case.
In the same paper we studied the relevant scales for structure formation, 
finding some interesting results:
1) the mirror Jeans mass $ M'_{\rm J} $ is always smaller than the ordinary 
one, thus making easier the growth of perturbations; 
2) there exist the dissipative mirror Silk scale $ M'_{\rm S} $ (analogous to 
the Silk scale for ordinary baryons), that for $ x \sim x_{\rm eq} $ has the 
value of a typical galaxy mass; 
3) for $ x < x_{\rm eq} $ we obtain $ M'_{\rm J} \sim M'_{\rm S} $, so that all the 
primordial perturbations with masses greater than the mirror Silk mass can 
grow uninterruptedly.

In the last decade the study of the the Cosmic Microwave Background (CMB) 
and the Large Scale Structure (LSS) is providing a great amount of 
observational data, and their continuous improvement provides powerful 
cosmological instruments that could help us to understand the nature of the 
dark matter of the Universe. 
In this context, the analysis of the CMB and LSS power spectra for a mirror 
baryonic dark matter scenario plays a key role as a cosmological test 
of the mirror theory. 
This is exactly the aim of this paper, in which we computed these spectra 
for a large range of parameters, qualitatively studied the consistence with 
present observational data, and obtained new useful limits on the mirror 
parameter space, broadening our previous studies \cite{paolo} with many 
new models and a detailed analysis of their dependence on all the 
cosmological parameters.

The plan of the paper is as follows. 
In next section we briefly describe the computed mirror models and the 
procedure used to obtain the CMB and LSS spectra.
In sections 3 and 4 we show respectively the CMB and LSS power 
spectra for a MBDM scenario, and study their dependence on the mirror 
parameters, comparing with the CDM case. 
In addition, we computed the spectrum of mirror CMB photons, 
comparing it with the analogous for the ordinary sector, and we extended 
the LSS spectra also to smaller non linear scales. 
Section 5 analyzes the dependence and the sensitivity of the CMB and 
LSS power spectra on the cosmological parameters; here we study different mirror models 
changing one parameter each time. 
Section 6 is devoted to the comparison of mirror power spectra with current 
observations, that allows us to obtain some bound on the mirror parameter 
space.
Finally, our main conclusions are summarized in section 7.


\section{The mirror models}
\label{mirror_mod}

We computed many models for Mirror Universe, assuming adiabatic scalar 
primordial perturbations, a flat space-time geometry, and different mixtures 
of ordinary and mirror baryons, photons and massless neutrinos, cold dark 
matter, and cosmological constant.

We used the evolutionary equations in synchronous gauge\footnote{
The difference in the use of other gauges is limited to the gauge-dependent 
behaviour of the density fluctuations on scales larger than the horizon. 
The fluctuations can appear as growing modes in one coordinate system 
and as constant mode in another, as shown in ref.~\cite{mabert} for the 
synchronous and the conformal Newtonian gauges.}
described in ref.~\cite{mabert}. 
Since we are working in the Fourier space $ k $, all the {\bf k} modes (where 
{\bf k} is the comoving wavevector and $ k $ its magnitude) in the linearized 
Einstein, Boltzmann, and fluid equations evolve independently; thus the 
equations can be solved for one value of {\bf k} at a time. 
Moreover, all modes with the same $ k $ obey the same evolutionary 
equations. 

In the same paper, the Boltzmann equations for massless neutrinos and 
photons have been transformed into an infinite hierarchy of moment 
equations $ F_l $ that must be truncated at some maximum multipole 
order $ l_{\rm max} $.  
Following the suggestions of authors, an improved truncation 
scheme\footnote{
One simple method is to set $ F_l = 0 $ for $ l > l_{\rm max} $, 
but it is inaccurate.
The authors affirmed that the problem with this scheme is the coupling of 
multipoles in equations, that leads to the propagation of errors from 
$ l_{\rm max} $ to smaller $ l $. 
These errors can propagate to $ l = 0 $ in a time 
$ \tau \approx l_{\rm max} / k $ and then reflect back to increasing $ l $, 
leading to amplification of errors in the interval
$ 0 \le l \le l_{\rm max} $ \cite{mabert}.} 
is based on extrapolating the behaviour of $ F_l $ to $ l = l_{\rm max} + 1 $ 
as
\begin{equation}
\label{truncnu}
F_{(l_{\rm max}+1)} \approx {(2l_{\rm max}+1) \over k \tau} \, 
  F_{l_{\rm max}} - F_{(l_{\rm max}-1)} \; ,
\end{equation}
where $ \tau $ is the conformal time.
However, time-variations of the potentials during the radiation-dominated 
era make even equation (\ref{truncnu}) a poor approximation if $ l_{\max} $ 
is chosen too small.
Thus, in order to obtain a relative accuracy better than $ 10 ^{-3} $ in our 
final results, in the computation of the potential and the density fields the 
photons and the massless neutrinos phase space distributions for both the 
ordinary and mirror sectors were expanded in Legendre series 
truncating the Boltzmann hierarchies at $ l_{\rm max} = 2000 $, using the 
truncation schemes given by the above expression (\ref{truncnu}) for 
massless neutrinos and equations (65) of ref. \cite{mabert} for photons.

We integrated in this way the equations of motion numerically over the range 
$ -5.0 \leq \log k \leq -0.5 $ (where $ k $ is measured in $ {\rm Mpc}^{-1} $) 
using points evenly spaced with an interval of $ \Delta \log k = 0.01 $. 
For purposes of computing the LSS power spectra for larger (non linear) 
scales (see \S~\ref{lss_non_lin}) we extended the integration of some 
models as far as $ \log k = 1.0 $. 
All the integrations were carried to the redshift $ z = 0 $.

The numerical computations were made using a Fortran code 
that solves the cited equations.
We took a pre-existent program written for a standard Universe, made only 
of the ordinary sector, and then we modified the code in order to simulate 
a Mirror Universe, which has a second sector. 
This required to modify and add several subroutines to the code. 
In fact we are dealing now with two self-interacting sectors, and thus all the 
equations governing the evolution of the considered components, 
relativistic (photons and massless neutrinos) and non relativistic (baryons), 
must be doubled. 
The cold dark matter equations were not doubled because they have the 
same physical effects independently of the sector.
In addition, the two sectors communicate via gravity, so that they are 
coupled and influence each other through this interaction. 
Therefore, there are more regimes than in the standard case, because now 
they are made up of the couplings between the different ordinary and mirror 
regimes, the latter being time-shifted from the former according to the laws 
exposed in previous section.

At last, we normalized our results to the COBE data, following the procedure 
described by Bunn and White in 1997 \cite{bunnwhite}.

\begin{table}[h]
\begin{center}
\begin{tabular}{|c|c|c|} \hline \hline
  Parameter & min. value & max. value \\
  \hline \hline
  $ \Omega_{\rm m} $ & 0.1 & 0.5 \\
  \hline
  $ \omega_{\rm b} $ & 0.010 & 0.030 \\
  \hline
  $ \omega'_{\rm b} $ & 0.0 & $ \omega_{\rm m} - \omega_{\rm b} $ \\
  \hline
  $ x $ & 0.1 & 0.7 \\
  \hline
  $ h $ & 0.5 & 0.9 \\
  \hline
  $ n $ & 0.90 & 1.10 \\
  \hline \hline
\end{tabular}
\end{center}
\caption{\small Parameters and their ranges used in mirror models. 
The values are not evenly spaced, but arbitrarily chosen in the parameter 
space. 
Not listed there are the total and vacuum densities, which, being flat 
models, are 
$ \Omega_0 = 1 $ and $ \Omega_\Lambda = 1 - \Omega_m $.}
\label{parrange}
 \end{table}

For the mirror models, we considered different values of the cosmological 
parameters, where we add to the usual ones two new mirror parameters: 
the ratio of the temperatures in the two sectors $ x = {T' / T} $ 
and the mirror baryons density $ \Omega'_b = \beta \Omega_b $ (also 
expressed via the ratio $ \beta $ of the baryonic densities in the two sectors).
Starting from an ordinary reference model, 
we study the influence of the mirror sector varying the two parameters that 
describe it for a given ordinary sector, replacing a fraction of CDM (or 
the entire dark matter) with MBDM. 
Furthermore, we evaluate the influence of the cosmological parameters for 
a mirror baryonic dark matter scenario, changing all of them. 
The values used for the parameters are not on a regular grid, but arbitrarily 
chosen for the only purpose to better understand the CMB and LSS for a 
Mirror Universe. 
At present we are not able to make a grid thin enough to perform a fit of 
all free parameters (even if we fix some of them) for two reasons: first, 
we have two further parameters, which lengthen a lot the computational 
time (by a factor order $ 10^2 $), and second our program is much slower 
than the ones commonly used for a standard Universe, because our models 
are more complicated in terms of calculus and for our choice to privilege 
precision instead of performance (at least at this stage\footnote{
The aim of this paper is not a fit of the parameters, but a study of some 
cosmological observables in a MBDM scenario, and possibly a qualitative 
test of the mirror theory. 
After this work, a future step could be just to write a new program 
much faster than the one here used, which should allow us to fit the 
parameters and compare the results with other cosmological models.}). 
We list the parameters used and their ranges in table \ref{parrange};
the total and vacuum densities (not listed in the table) are fixed by our 
choice of a flat geometry: $ \Omega_0 = 1 $ 
and $ \Omega_\Lambda = 1 - \Omega_m $. 
In addition, in order to study the parameter dependence, we computed 
models also for different numbers of extra-neutrino species 
($ \Delta N_\nu =0.5, 1.0, 1.5 $).


\section{The cosmic microwave background for a Mirror Universe}
\label{cmb_2}

\begin{figure}[p]
  \begin{center}
    \leavevmode
    \epsfxsize =12cm
    \epsffile{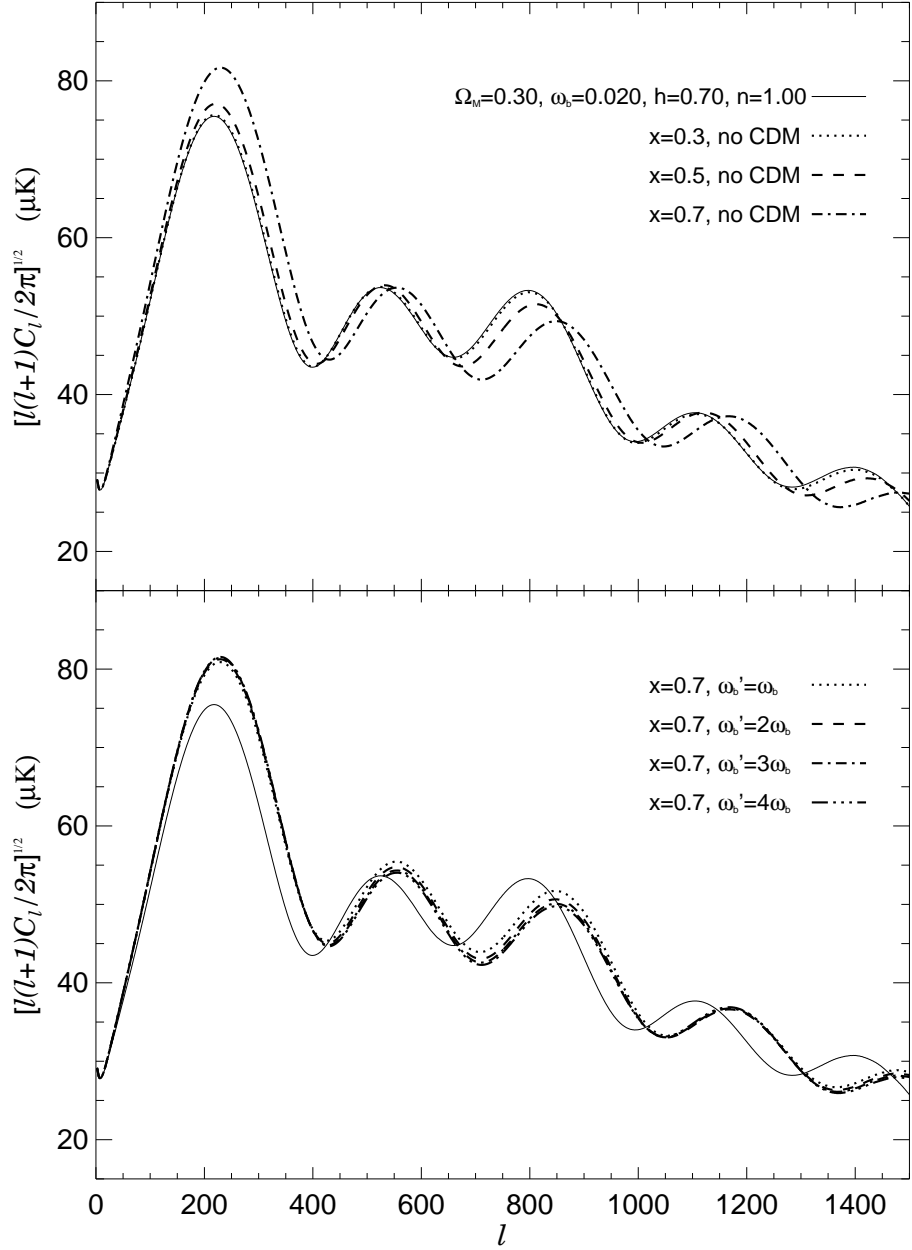}
  \end{center}
\vspace{-.4cm}
\caption{\small CMB angular power spectrum for different values of $ x $ 
and $ \omega_b' = \Omega_b' h^2 $, compared with a standard 
model (solid line). 
{\sl Top panel.} Mirror models with the same parameters as the ordinary 
one, and with $ x = 0.3, 0.5, 0.7 $, and 
$ \omega_b' = \Omega_m h^2 - \omega_b $ (no CDM) for all 
models.
{\sl Bottom panel.} Mirror models with the same parameters as the ordinary 
one, and with $ x = 0.7 $ and 
$ \omega_b' = \omega_b, 2 \omega_b, 3 \omega_b, 4 \omega_b $.}
\label{cmblssfig1}
\end{figure}

As anticipated when we studied the structure formation in Paper I 
\cite{paper1} (and predicted or shown in refs.~\cite{ignavol-lss,paolo,bcv}), 
we expect that the existence of a mirror sector influences the 
cosmic microwave background radiation observable today; in this section 
we want to evaluate this effect.

We choose a starting standard model and add a mirror sector simply 
removing cold dark matter and adding mirror baryons. 
The values of the parameters for this reference model are: 
$ \Omega_0 = 1 $, $ \Omega_m = 0.30 $, $ \Omega_\Lambda = 0.70 $, 
$ \omega_b = \Omega_b h^2 = 0.02 $, $ n = 1.00 $, $ h = 0.70 $, 
with only cold dark matter (no massive neutrinos) and scalar 
adiabatic perturbations (with spectral index $ n $).
This reference model is not the result of a fit, but is arbitrarily chosen 
consistent with the current knowledge of the cosmological parameters; 
however, this is not a shortcoming, because here we want only to put in 
evidence the differences from a representative reference model (without 
comparison with observations), and this is a good model for this intention.

From this starting point, first of all we substitute all the cold dark matter with 
mirror baryonic dark matter (MBDM) and evaluate the CMB angular power 
spectrum varying $ x $ from 0.3 to 0.7 (around the upper limit set by the 
BBN bounds). 
This is shown in top panel of figure \ref{cmblssfig1}, where mirror models 
are plotted together with the reference model. 
The first evidence is that the deviation from the standard model is not linear 
in $ x $: it grows more for bigger $ x $ and for $ x \lsim 0.3 $ the power 
spectra are practically coincident. 
This is important, because it means that a Universe where all the dark 
matter is made of mirror baryons could be indistinguishable from a CDM 
model if we analyze the CMB only. 
We see the greatest separation from the reference model for $ x = 0.7 $, but 
it will increase for hypothetical larger values of $ x $. 
The height of the first acoustic peak grows for $ x \gsim 0.3 $, while the 
position remains nearly constant. 
For the second peak the opposite occurs, i.e. the height remains practically 
constant, while the position shifts toward higher multipoles $ l $; for the 
third peak, instead, we have a shift both in height and position (the absolute 
shifts are similar to the ones for the first two peaks, but the height now 
decreases instead of increasing).
Observing also other peaks, we recognize a general pattern: except for 
the first one, odd peaks change both height and location, even ones change 
location only.

In bottom panel of figure \ref{cmblssfig1} we show the intermediate case of 
a mixture of CDM and MBDM. 
We consider  $ x = 0.7$, a high value which permits us to see well the 
differences changing $ \omega_b' $ from  $ \omega_b $ to $ 4 \omega_b $. 
The dependence on the amount of mirror baryons is lower than on the 
ratio of temperatures $ x $. 
In fact, the position of the first peak is nearly stable for all mirror models 
(except for a very low increase of height for growing 
$ \omega_b' $), while differences appear for other peaks. 
In the second peak the position is shifted as in the case without CDM 
independently of $ \omega_b' $, while the height is inversely 
proportional to $ \omega_b' $ with a separation appreciable for 
$ \omega_b' \lsim 3 \omega_b $. 
For the third peak the behaviour is the same as for the case without CDM, 
with a slightly stronger dependence on $ \omega_b' $, while for the 
other peaks there is a weaker dependence on $ \omega_b' $. 
A common feature is that the heights of the peaks are not linearly 
dependent on the mirror baryonic density, while their positions are 
practically insensitive to $ \omega_b' $ but depend only on $ x $.

We will analyze in more detail the $ x $ and $ \omega_b' $ 
dependence of the peaks, together with other parameters, in 
\S~\ref{para_trends}.


\subsection{The mirror cosmic microwave background radiation}
\label{cmb_3}

In the same way as ordinary photons at decoupling from baryons formed 
the CMB we observe today, also mirror photons at their decoupling formed 
a mirror cosmic microwave background radiation, which, unfortunately, we 
cannot observe because they don't couple with the ordinary baryons of 
which we are made\footnote{
Indeed, there is in principle the possibility that mirror CMB photons could 
influence our CMB photons in case of existence of a photon-mirror photon 
kinetic mixing \cite{mixing}, 
but its detection would not be possible with present and probably even 
future experiments, given the very low estimates for the cross 
section of this interaction.
} (instead, it would be possible for an hypothetical mirror observer\footnote{
Since two worlds have the same microphysics, the life should be 
possible also in the mirror sector!}). 
Nevertheless, its study is not only speculative, since it is a way to better 
understand the cosmology with MBDM and our observable CMB.

\begin{figure}[h]
  \begin{center}
    \leavevmode
    \epsfxsize = 12cm 
    \epsffile{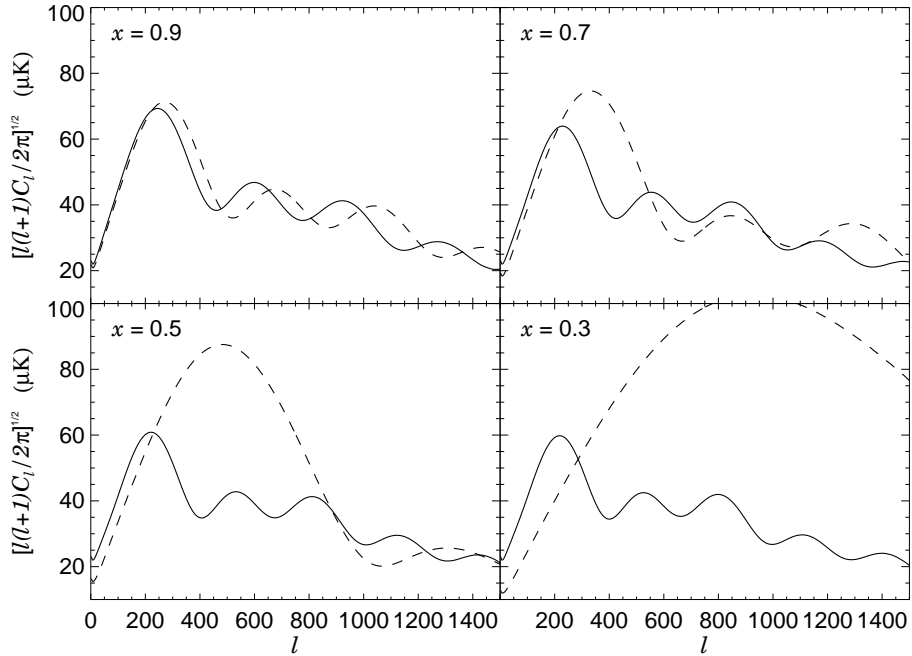}
  \end{center}
\caption{\small Angular power spectra for ordinary (solid line) and mirror 
(dashed line) CMB photons. The models have $ \Omega_0 = 1 $, 
$ \Omega_m = 0.3 $, $ \omega_b = \omega_b' = 0.02 $, $ h = 0.7 $, 
$ n = 1.0 $, and $ x = 0.9 $ (top-left panel), $ x = 0.7 $ (top-right panel), 
$ x = 0.5 $ (bottom-left panel)$, x = 0.3 $ (bottom-right panel).}
\label{cmblssfig2}
\end{figure}

We computed four models of mirror CMB, in order to have enough elements 
to compare with the corresponding observable CMBs. 
The chosen parameter values are those usually taken, and the amount of 
mirror baryons is the same as the ordinary ones, while $ x $ is taken as 
0.9, 0.7, 0.5 or 0.3 in order to explore different scenarios. 
Thus the parameters of the models are: $ \Omega_0 = 1 $, 
$ \Omega_m = 0.3 $, $ \omega_b = \omega_b' = 0.02 $, 
$ x = 0.7$ or 0.5, $ h = 0.7 $, $ n = 1.0 $.
In figure \ref{cmblssfig2} we plot the ordinary and mirror CMB spectra 
corresponding to the same model of Mirror Universe. 

The first evidence is that, being scaled by the factor $ x $ the temperatures 
in the two sectors, also their temperature fluctuations will be scaled by the 
same amount, as evident if we look at the lowest $ \ell $ values (the 
fluctuations seeds are the same for both sectors). 
Starting from the top-left panel of the figure, we see that the first mirror CMB 
peak is higher and shifted to higher multipoles than the ordinary one, 
while other peaks are both lower and at higher $ \ell $ values, 
with a shift growing with the order of the peak. 

Observing all the panels, we understand the effect of a change of the 
parameter $ x $ on the mirror CMB: 
(i) for lower $ x $-values the first peak is higher (for $ x = 0.5 $ it is nearly 
one and half the ordinary one); 
(ii) the position shifts to much higher multipoles (so that with the same 
horizontal scale we can no more see some peaks). 
The reason is that a change of $ x $ corresponds to a change of the mirror 
decoupling time. 
The mirror photons, which decouple before the ordinary ones, see a smaller 
sound horizon, scaled approximately by the factor $ x $; since the first peak 
occurs at a multipole $ \ell \propto ({\rm sound\;horizon}) ^{-1} $, we expect it 
to shifts to higher $ \ell $-values by a factor $ x^{-1} $, that is exactly what 
we observe in the figure.

We have verified that increasing $ x $ the 
mirror CMB is more and more similar to the ordinary one, until for $ x = 1 $ 
(not shown in figure) the two power spectra are perfectly coincident 
(as expected, since in this 
case the two sectors have exactly the same temperatures, the same particle 
contents, and then their photons power spectra are necessarily the same). 

If we were able to detect both the ordinary and mirror CMB photons, we had 
two snapshots of the Universe at two different epochs, which were a 
powerful cosmological instrument, but unfortunately this is impossible, 
because mirror photons are by definition completely invisible for us.


\section{The large scale structure for a Mirror Universe}
\label{lss_2}

Given the oscillatory behaviour of the mirror baryons (different from the 
smooth one of cold dark matter), we expect that MBDM induces specific 
signatures also on the large scale structure power spectrum. 

\begin{figure}[p]
  \begin{center}
    \leavevmode
    \epsfxsize = 12cm
    \epsffile{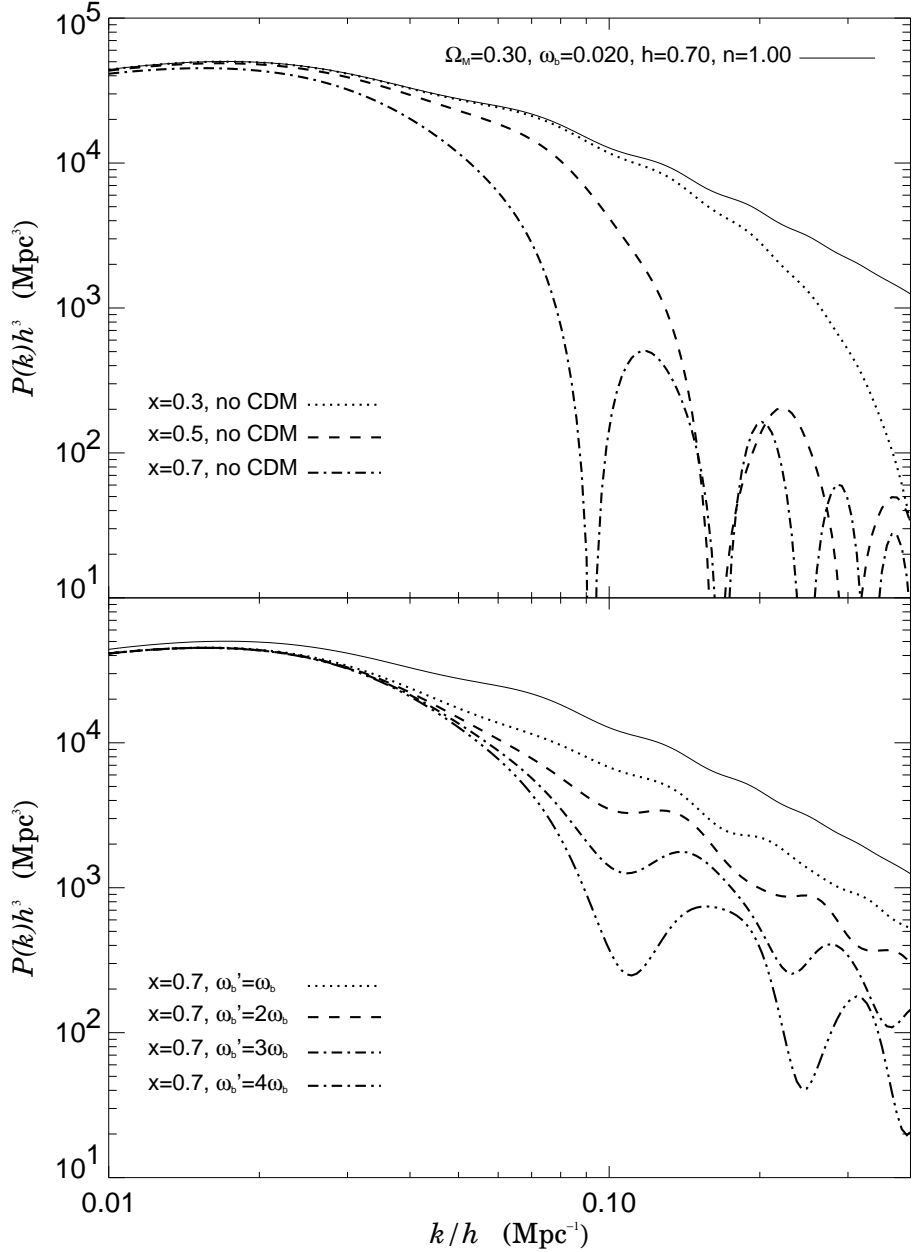}
  \end{center}
\addvspace{.8cm}
\caption{\small LSS power spectrum in the linear regime for different values 
of $ x $ and $ \omega_b' = \Omega_b' h^2 $, compared with a 
standard model (solid line). 
In order to remove the dependences of units on the Hubble constant, we plot 
on the $ x $-axis the wave number in units of $ h $ and on the $ y $-axis the 
power spectrum in units of $ h^3 $.
{\sl Top panel.} Mirror models with the same parameters as the ordinary 
one, and with $ x = 0.3, 0.5, 0.7 $ and 
$ \omega_b' = \Omega_m h^2 - \omega_b $ (no CDM) for all 
models.
{\sl Bottom panel.} Mirror models with the same parameters as the ordinary 
one, and with $ x = 0.7 $ and 
$ \omega_b' = \omega_b, 2 \omega_b, 3 \omega_b, 4 \omega_b $.}
\label{cmblssfig3}
\end{figure}

In order to evaluate this effect, we computed LSS power spectra using the 
same reference and mirror models used in \S~\ref{cmb_2} for the CMB 
analysis. 
The two panels of figure \ref{cmblssfig3} show the LSS for the same models 
as in figure \ref{cmblssfig1}. 
In order to remove the dependences of units on the Hubble constant, we 
plot on the $ x $-axis the wave number in units of $ h $ and on the 
$ y $-axis the power spectrum in units of $ h^3 $. 
The minimum scale (the maximum $ k $) plotted is placed around the limit 
of the linear regime.

In top panel of the figure we show the dependence on $ x $ for different 
mirror models without CDM; in this case, where all the dark matter is made 
of mirror baryons, the oscillatory effect is obviously maximum. 
The first evidence is the strong dependence on $ x $ of the beginning of 
oscillations: it goes to higher scales for higher $ x $, and below 
$ x \simeq 0.3 $ the power spectrum for a Mirror Universe approaches 
more and more the CDM one. 
This behaviour is a consequence of the $ x $-dependence of the mirror Silk 
scale, that increases for growing $ x $ (for details see Paper I \cite{paper1}
and refs.~\cite{ignavol-lss,paolo,bcv}): 
this dissipative scale induces a cutoff in the power spectrum, which is 
damped with an oscillatory behaviour (it will be more evident in figures 
\ref{cmblssfig4} and \ref{cmblssfig5}, where we extend our models to 
smaller scales within the non linear region). 
Oscillations begin at the same time of the damping, and they are so deep 
(because there are many mirror baryons) to go outside the coordinate box. 
In any case the mirror spectra are always below the ordinary one for every 
value of $ x $.

The dependence on the amount of mirror baryons is instead shown in the 
bottom panel of the figure, where only a fraction of the dark matter is 
made of mirror baryons, while the rest is CDM. 
Contrary to the CMB case, the matter power spectrum strongly depends on 
$ \omega_b' $. 
The oscillations are deeper for increasing mirror baryon densities and the 
spectrum goes more and more away from the pure CDM one. 
We note also that the damping begins always at the same scale, and thus it 
depends only on $ x $ and not on $ \omega_b' $, as we know from 
the expression of the mirror Silk scale obtained in Paper I \cite{paper1} 
and refs.~\cite{ignavol-lss,paolo,bcv} 
\be \label{ms_m}
 M'_S \sim [f(x) / 2]^3 (\omega_b')^{-5/4} 10^{12}~ M_\odot \; , 
\ee
where $ f(x) = x^{5/4} $ for $ x > x_{\rm eq} $ and 
$ f(x) = (x / x_{\rm eq})^{3/2} x_{\rm eq}^{5/4} $ for $ x < x_{\rm eq} $. 

The same considerations are valid for the oscillation minima, which 
become much deeper for higher mirror baryon densities, but shift very 
slightly to lower scales, so that their positions remain practically constant.


\subsection{Extension to smaller (non linear) scales}
\label{lss_non_lin}

\begin{figure}[p]
  \begin{center}
    \leavevmode
    \epsfxsize = 12cm
    \epsffile{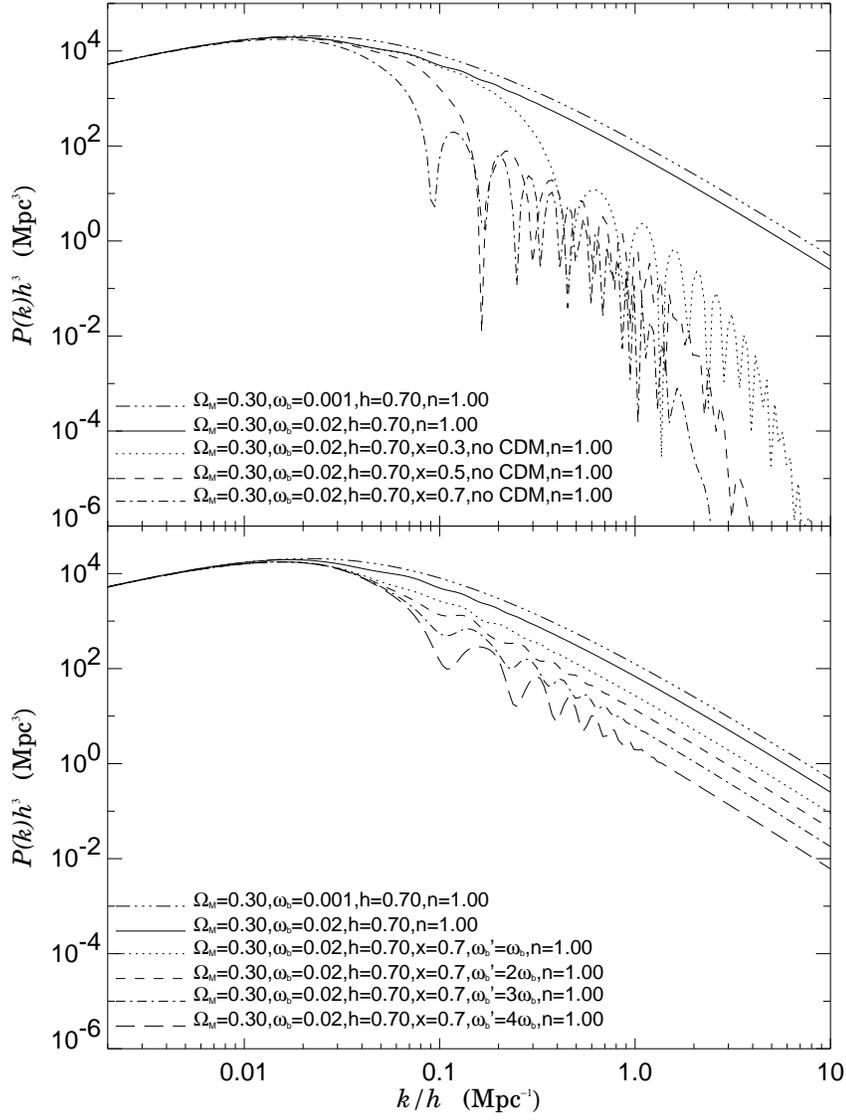}
  \end{center}
\addvspace{.8cm}
\caption{\small LSS power spectrum beyond the linear regime for different 
values of $ x $ and $ \omega_b' = \Omega_b' h^2 $, compared 
with a standard model (solid line). 
The models have the same parameters as in figure \ref{cmblssfig3}.
For comparison we also show a standard CDM model with a negligible 
amount of baryons ($ \Omega_b \sim 0.2 \% $).}
\label{cmblssfig4}
\end{figure}

Let us now extend the behaviour of the matter power spectrum to lower 
scales, which already became non linear. 
Obviously, since our treatment is based on the linear theory, it is no longer 
valid in non linear regime. 
Nevertheless, even if it cannot be used for a comparison with observations, 
the extension of our models to these scales is very useful to understand the 
behaviour of the power spectrum in a mirror baryonic dark matter scenario, 
in particular concerning the position of the cutoff (we recall that its presence 
could help in avoiding the problem of the CDM scenario with the excessive 
number of small structures).

\begin{figure}[h]
  \begin{center}
    \leavevmode
    \epsfxsize = 12cm
    \epsffile{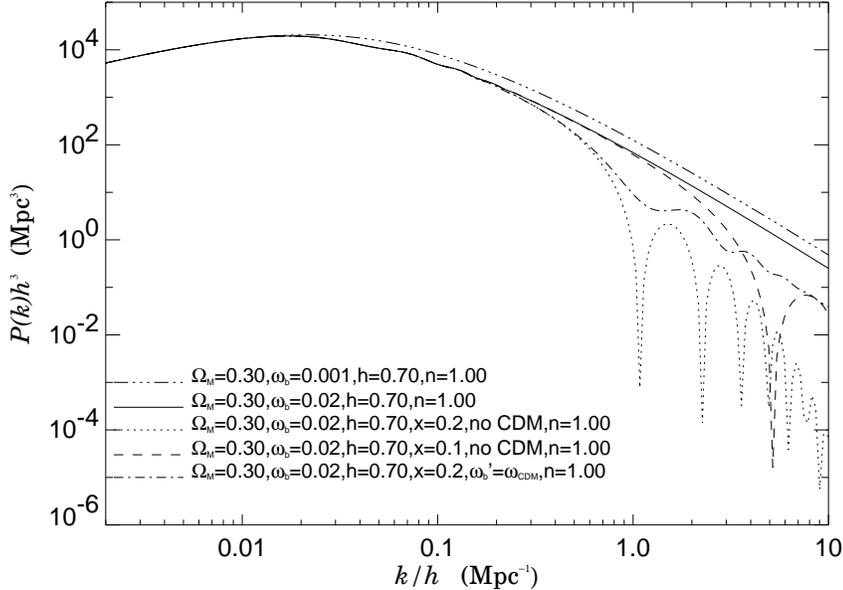}
  \end{center}
\addvspace{-6.5cm}
\caption{\small LSS power spectrum beyond the linear regime for two low 
values of $ x $ (0.1 and 0.2) and different amounts of mirror baryons 
($ \Omega_b' = \Omega_m - \Omega_b $ or 
$ \Omega_b' = \Omega_{CDM} $), compared with a standard model 
(solid line). 
The other parameters are the same as in figure \ref{cmblssfig3}.
For comparison we show also a standard CDM model with a negligible 
amount of baryons ($ \Omega_b \sim 0.2\% $).}
\label{cmblssfig5}
\end{figure}

Therefore, in figures \ref{cmblssfig4} and \ref{cmblssfig5} we extend the 
power spectra up to $ k/h = 10 $ Mpc$^{-1}$ (corresponding to galactic 
scales), well beyond the limit of the linear regime, given approximately by 
$ k/h < 0.4 $ Mpc$^{-1}$.

In figure \ref{cmblssfig4} we plot in both panels the same models as in 
figure \ref{cmblssfig3}.
For comparison we show also a standard model characterized by a matter 
density made almost completely of CDM, with only a small contamination of 
baryons ($ \Omega_b \simeq 0.2\% $ instead of $ \simeq 4 \% $ of 
other models). 
In the top panel, the $ x $-dependence of the mirror power spectra is 
considered: the vertical scale extends to much lower values compared to 
figure \ref{cmblssfig3}, and we can clearly see the deep oscillations, but in 
particular it is evident the presence of the previously cited cutoff. 
For larger values of $ x $ oscillations begin earlier and cutoff moves to 
higher scales. 
Moreover, note that the model with almost all CDM has more power than the 
same standard model with baryons, which in turn has more power than all 
mirror models for any $ x $ and for all the scales. 
In the bottom panel we show the dependence on the mirror baryon content. 
It is remarkable that all mirror models stop to oscillate at some 
low scale and then continue with a smooth CDM-like trend. 
This means that, after the cutoff due to mirror baryons, the dominant 
behaviour is the one characteristic of cold dark matter models (due to the 
lack of a cutoff for CDM). 
Clearly, for higher mirror baryon densities the oscillations continue down to 
smaller scales, but, contrary to the previous case, where all the dark matter 
was mirror baryonic, there will always be a scale below which the spectrum 
is CDM-like.

An interesting point of the mirror baryonic scenario is his capability to mimic 
a CDM scenario under certain circumstances and for certain measurements. 
To explain this point, in figure \ref{cmblssfig5} we show models with low 
$ x $-values (0.2 or 0.1) and all dark matter made of MBDM; we see that 
for $ x = 0.2 $ the standard and mirror power spectra are already practically 
coincident in the linear region. 
If we go down to $ x = 0.1 $ the coincidence is extended up to $ k/h \sim 1 $ 
Mpc$ ^{-1} $. 
In principle, we could still decrease $ x $ and lengthen this region of 
equivalence between the different CDM and MBDM models, but we have 
to remember that we are dealing with linear models 
extended to non linear scales, then neglecting all the non linear phenomena 
(such as merging or stellar feedback), that are very different for the CDM and 
the MBDM scenarios. 
In the same plot we also considered a model with $ x = 0.2 $ and dark 
matter composed equally by mirror baryons and CDM. 
This model shows that in principle it's possible a tuning of the cutoff 
effect reducing the amount of mirror matter, in order to better reproduce the 
cutoff needed to explain, for example, the low number of small satellites in 
galaxies.

This work provided for us the linear transfer functions, which constitute the 
principal ingredient for the computation of the power spectrum at non linear 
scales. 
This calculation is out of the aim of this paper, but could be one of the next 
steps in the study of the Mirror Universe.

\section{Dependence on the parameters}
\label{para_trends}

The shapes, heights and locations of peaks and oscillations in the photons 
and matter power spectra are predicted by all models based on the 
inflationary scenario. 
Furthermore the details of the features in these power spectra depend 
critically on the chosen cosmological parameters and on the composition of 
the Universe, which in turn can be accurately determined by precise 
measurements of these patterns. In this section we briefly discuss the 
sensitivity of the $ C_\ell $'s and $ P(k) $'s on the values of some 
fundamental parameters in a mirror baryonic scenario.

In particular, the exact form of the CMB and LSS power spectra is greatly 
dependent on assumptions about the matter content of the Universe. 
Apart from the total density parameter $ \Omega_0 $ (our models are flat, 
so $ \Omega_0 $ is always 1), the composition of the Universe can be 
parametrized by its components $ \Omega_m $ and 
$ \Omega_\Lambda $, and the components of the matter density 
$ \Omega_b$, $ \Omega_b' $, $ \Omega_{CDM} $. 
Further parameters are the tilt of scalar fluctuations $ n $, the Hubble 
parameter {\it h}, and the ratio of temperatures in the two sectors $ x $. 
In addition, we consider also the dependence on the number of massless 
neutrino species $ N_\nu $, in order to compare it with the 
$ x $-dependence. 
This is important if we remember that the relativistic mirror particles can be 
parametrized in terms of effective number of extra-neutrino species 
(for details see ref.~\cite{bcv}).

\begin{figure}[p]
  \begin{center}
    \leavevmode
    \epsfxsize = 12cm
    \epsffile{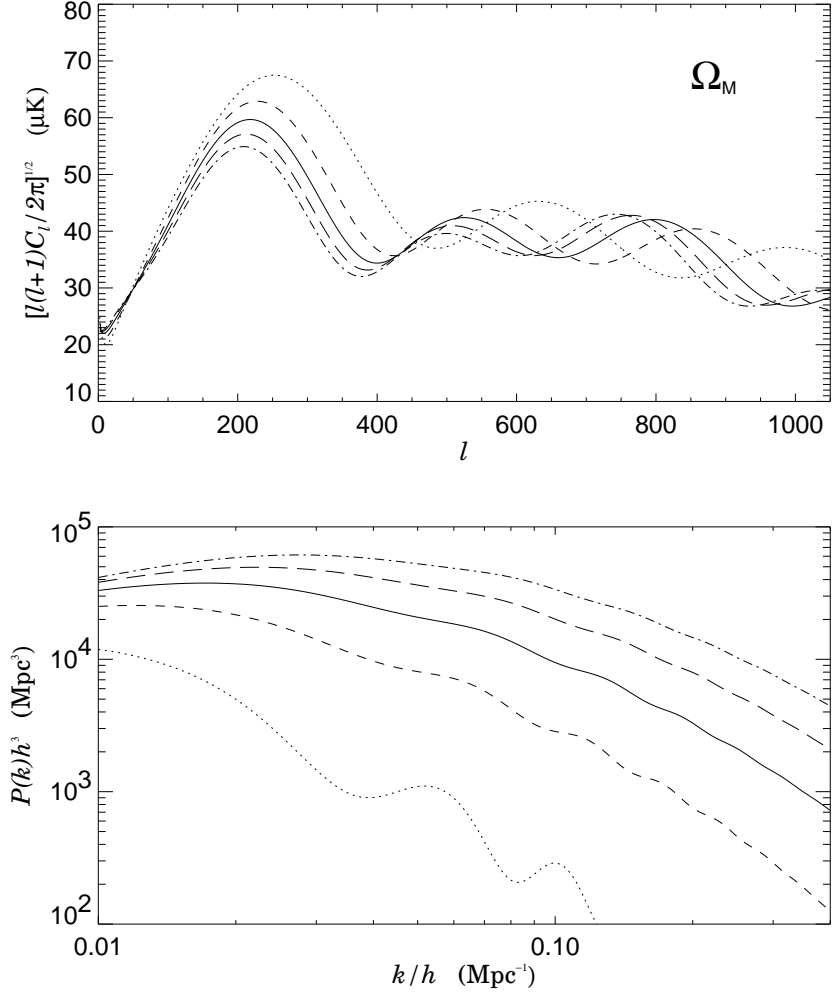}
  \end{center}
\caption{\small Dependence of the shape of photon and matter power 
spectra on the matter density $ \Omega_m $. 
The reference model (solid line) has: $ \Omega_m = 0.3 $, 
$ \omega_b = \omega_b' = 0.02 $, $ x = 0.2 $, $ h = 0.7 $ and 
$ n = 1.0 $. 
For other models, all the parameters are unchanged except for the one 
indicated: $ \Omega_m = 0.1 $ (dot), $ \Omega_m = 0.2 $ (dash), 
$ \Omega_m = 0.4 $ (long dash), $ \Omega_m = 0.5 $ (dot-dash).}
\label{cmblssfig8}
\end{figure}

\begin{figure}[p]
  \begin{center}
    \leavevmode
    \epsfxsize = 12cm
    \epsffile{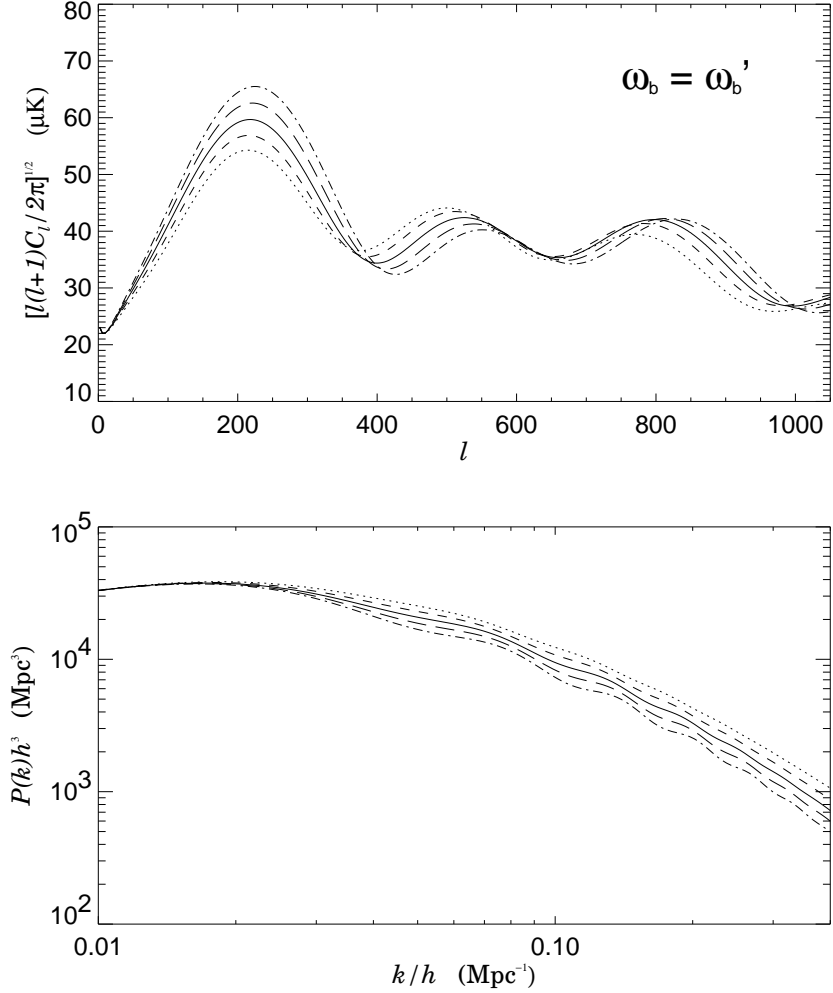}
  \end{center}
\caption{\small Dependence of the shape of photon and matter power 
spectra on the ordinary baryon density $ \omega_b $ with 
$ \omega_b = \omega_b' $. 
The reference model (solid line) has: $ \Omega_m = 0.3 $, 
$ \omega_b = \omega_b' = 0.02 $, $ x = 0.2 $, $ h = 0.7 $ and 
$ n = 1.0 $. 
For other models, all the parameters are unchanged except for the one 
indicated: $ \omega_b = 0.010 $ (dot), $ \omega_b = 0.015 $ (dash), 
$ \omega_b = 0.025 $ (long dash), $ \omega_b = 0.03 $ (dot-dash).}
\label{cmblssfig10}
\end{figure}

\begin{figure}[p]
  \begin{center}
    \leavevmode
    \epsfxsize = 12cm
    \epsffile{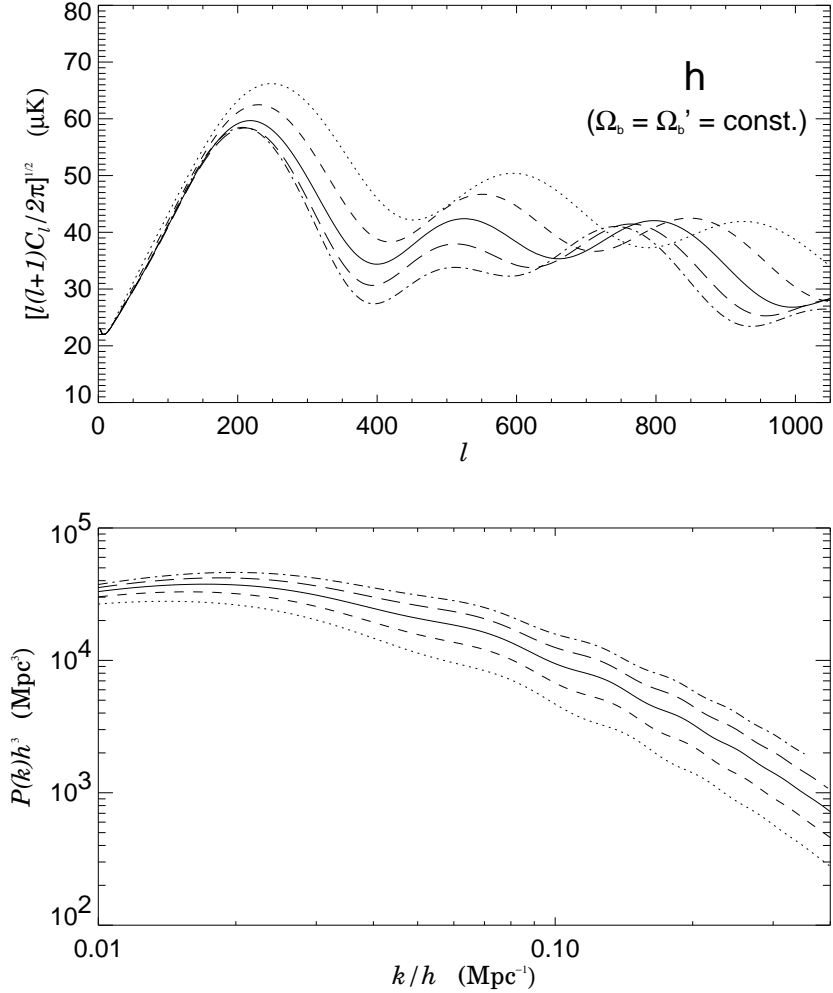}
  \end{center}
\caption{\small Dependence of the shape of photon and matter power 
spectra on the Hubble parameter $ h $ with 
$ \Omega_b = \Omega_b' = {\rm const} $. 
The reference model (solid line) has: $ \Omega_{\rm m} = 0.3 $, 
$ \Omega_b = \Omega_b' = 0.0408 $ (the value obtained for 
$ \omega_b = 0.02 $ and $ h = 0.7 $), $ x = 0.2 $, $ h = 0.7 $ and 
$ n = 1.0 $. 
For other models, all the parameters are unchanged except for the one 
indicated: $ h = 0.5 $ (dot), $ h = 0.6 $ (dash), $ h = 0.8 $ (long dash), 
$ h = 0.9 $ (dot-dash).}
\label{cmblssfig11}
\end{figure}

\begin{figure}[p]
  \begin{center}
    \leavevmode
    \epsfxsize = 12cm
    \epsffile{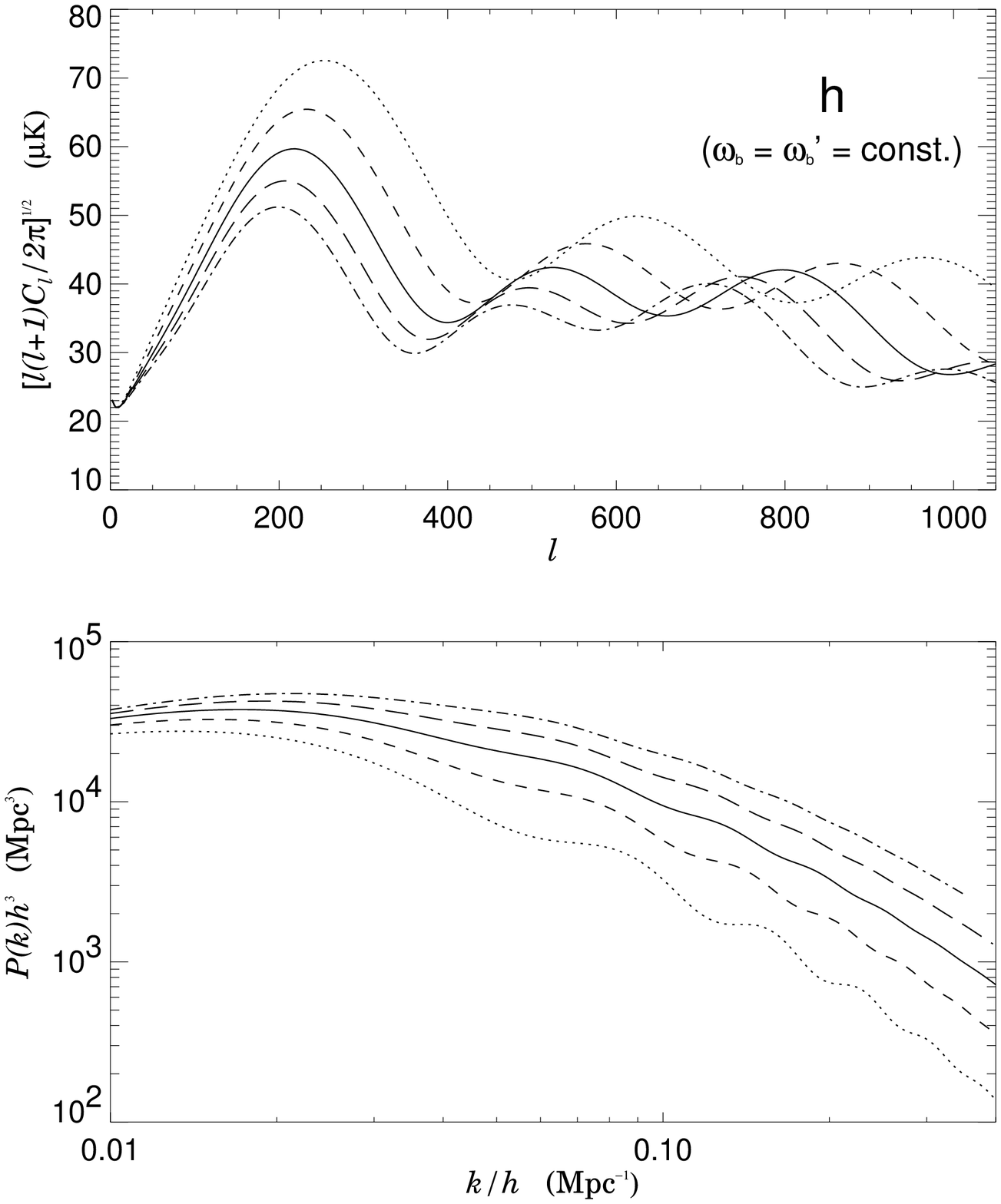}
  \end{center}
\caption{\small Dependence of the shape of photon and matter power 
spectra on the Hubble parameter $ h $ with 
$ \omega_b = \omega_b' = {\rm const} $. 
The reference model (solid line) has: $ \Omega_m = 0.3 $, 
$ \omega_b = \omega_b' = 0.02 $, $ x = 0.2 $, $ h = 0.7 $ and $ n = 1.0 $. 
For other models, all the parameters are unchanged except for the one 
indicated: $ h = 0.5 $ (dot line), $ h = 0.6 $ (dash), $ h = 0.8 $ (long dash), 
$ h = 0.9 $ (dot-dash).}
\label{cmblssfig12}
\end{figure}

\begin{figure}[p]
  \begin{center}
    \leavevmode
    \epsfxsize = 12cm
    \epsffile{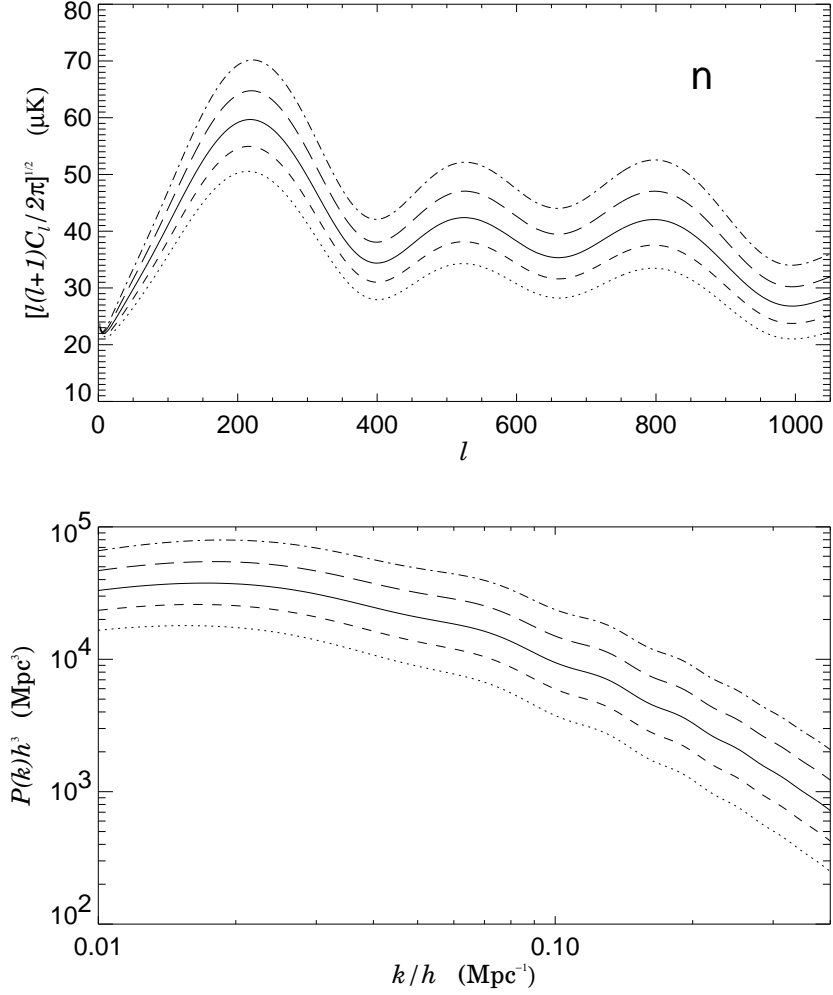}
  \end{center}
\caption{\small Dependence of the shape of photon and matter power 
spectra on the scalar spectral index $ n $. 
The reference model (solid line) has: $ \Omega_m = 0.3 $, 
$ \omega_b = \omega_b' = 0.02 $, $ x = 0.2 $, $ h = 0.7 $ and 
$ n = 1.0 $. 
For other models, all the parameters are unchanged except for the one 
indicated: $ n = 0.90 $ (dot), $ n = 0.95 $ (dash), $ n = 1.05 $ (long dash), 
$ n = 1.10 $ (dot-dash).}
\label{cmblssfig13}
\end{figure}

\begin{figure}[p]
  \begin{center}
    \leavevmode
    \epsfxsize = 12cm
    \epsffile{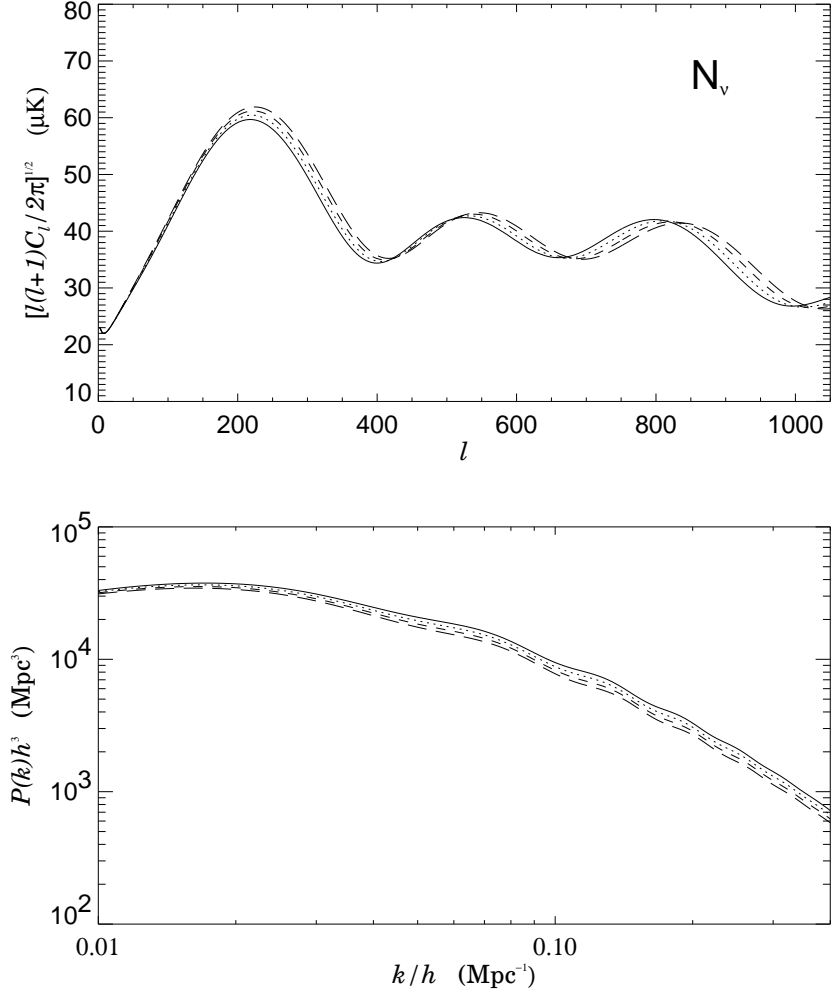}
  \end{center}
\caption{\small Dependence of the shape of photon and matter power 
spectra on the number of extra-neutrino species $ \Delta N_\nu $. 
The reference model (solid line) has: $ \Omega_m = 0.3 $, 
$ \omega_b = \omega_b' = 0.02 $, $ x = 0.2 $, $ h = 0.7 $, 
$ n = 1.0 $, and $ \Delta N_\nu = 0 $. 
For other models, all the parameters are unchanged except for the one 
indicated: $ \Delta N_\nu = 0.5 $ (dot), $ \Delta N_\nu = 1.0 $ (dash), 
$ \Delta N_\nu = 1.5 $ (long dash).}
\label{cmblssfig14}
\end{figure}

\begin{figure}[p]
  \begin{center}
    \leavevmode
    \epsfxsize = 12cm
    \epsffile{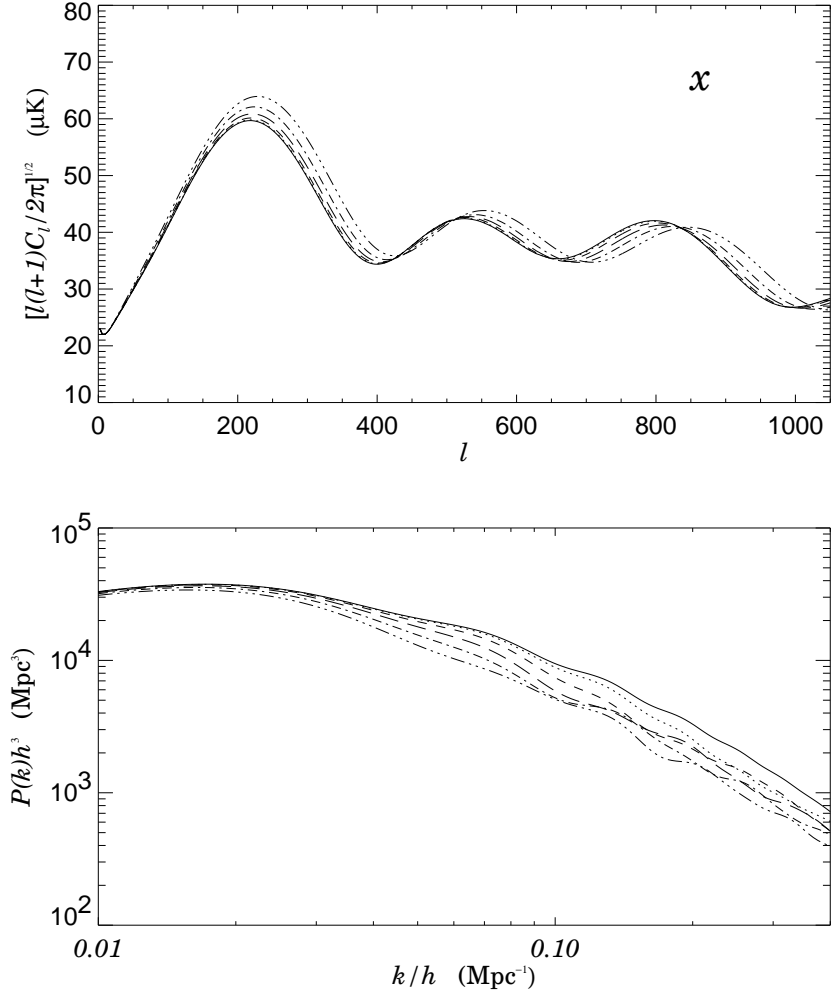}
  \end{center}
\caption{\small Dependence of the shape of photon and matter power 
spectra on the ratio of the temperatures of two sectors $ x $. 
The reference model (solid line) has: $ \Omega_m = 0.3 $, 
$ \omega_b = \omega_b' = 0.02 $, $ x = 0.2 $, $ h = 0.7 $ and 
$ n = 1.0 $. 
For other models, all the parameters are unchanged except for the one 
indicated: $ x = 0.3 $ (dot), $ x = 0.4 $ (dash), $ x = 0.5 $ (long dash), 
$ x = 0.6 $ (dot-dash), $ x = 0.7 $ (dot-dot-dot-dash).}
\label{cmblssfig15}
\end{figure}

\begin{figure}[p]
  \begin{center}
    \leavevmode
    \epsfxsize = 12cm
    \epsffile{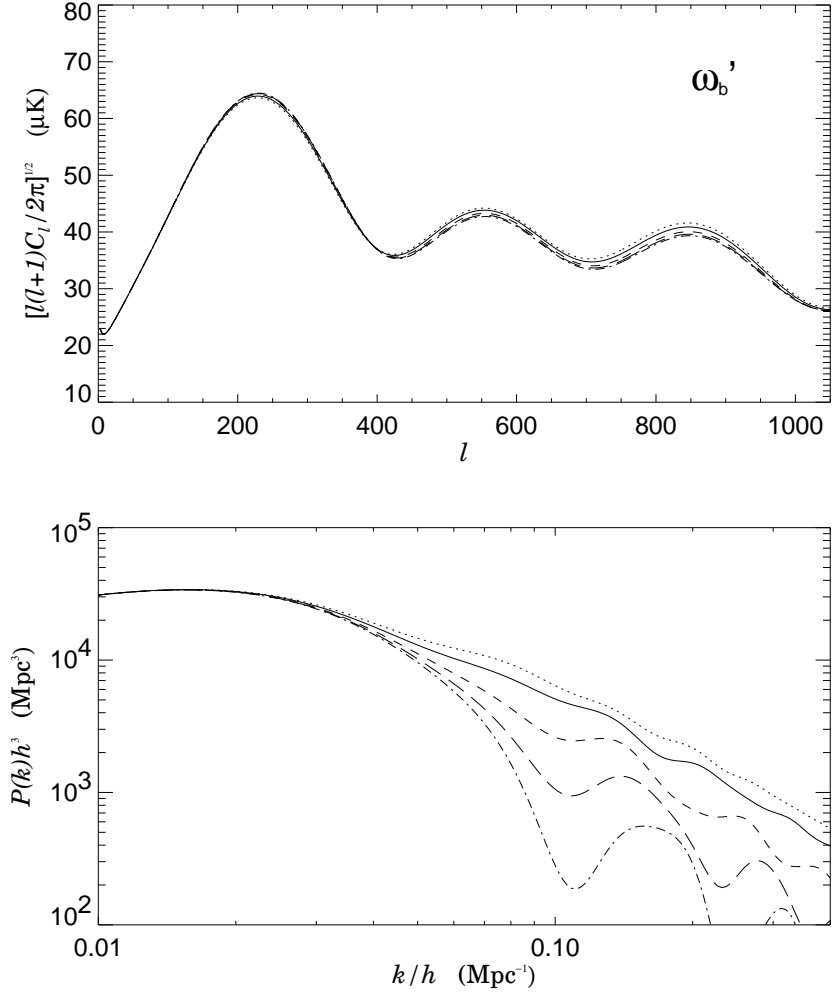}
  \end{center}
\caption{\small Dependence of the shape of photon and matter power 
spectra on the mirror baryon density $ \omega_b' $ keeping constant 
$ \omega_b $. 
The reference model (solid line) has: $ \Omega_m = 0.3 $, 
$ \omega_b = \omega_b' = 0.02 $, $ x = 0.7 $ (not 0.2, as 
previous figures), $ h = 0.7 $ and $ n = 1.0 $. 
For other models, all the parameters are unchanged except for the one 
indicated: $ \omega_b' = 0.01 = \omega_b / 2 $ (dot), 
$ \omega_b' = 0.04 = 2 \omega_b $ (dash), 
$ \omega_b' = 0.06 = 3 \omega_b $ (long dash), 
$ \omega_b' = 0.08 = 4 \omega_b$ (dot-dash).}
\label{cmblssfig17}
\end{figure}

\begin{figure}[p]
  \begin{center}
    \leavevmode
    \epsfxsize = 15cm
    \epsffile{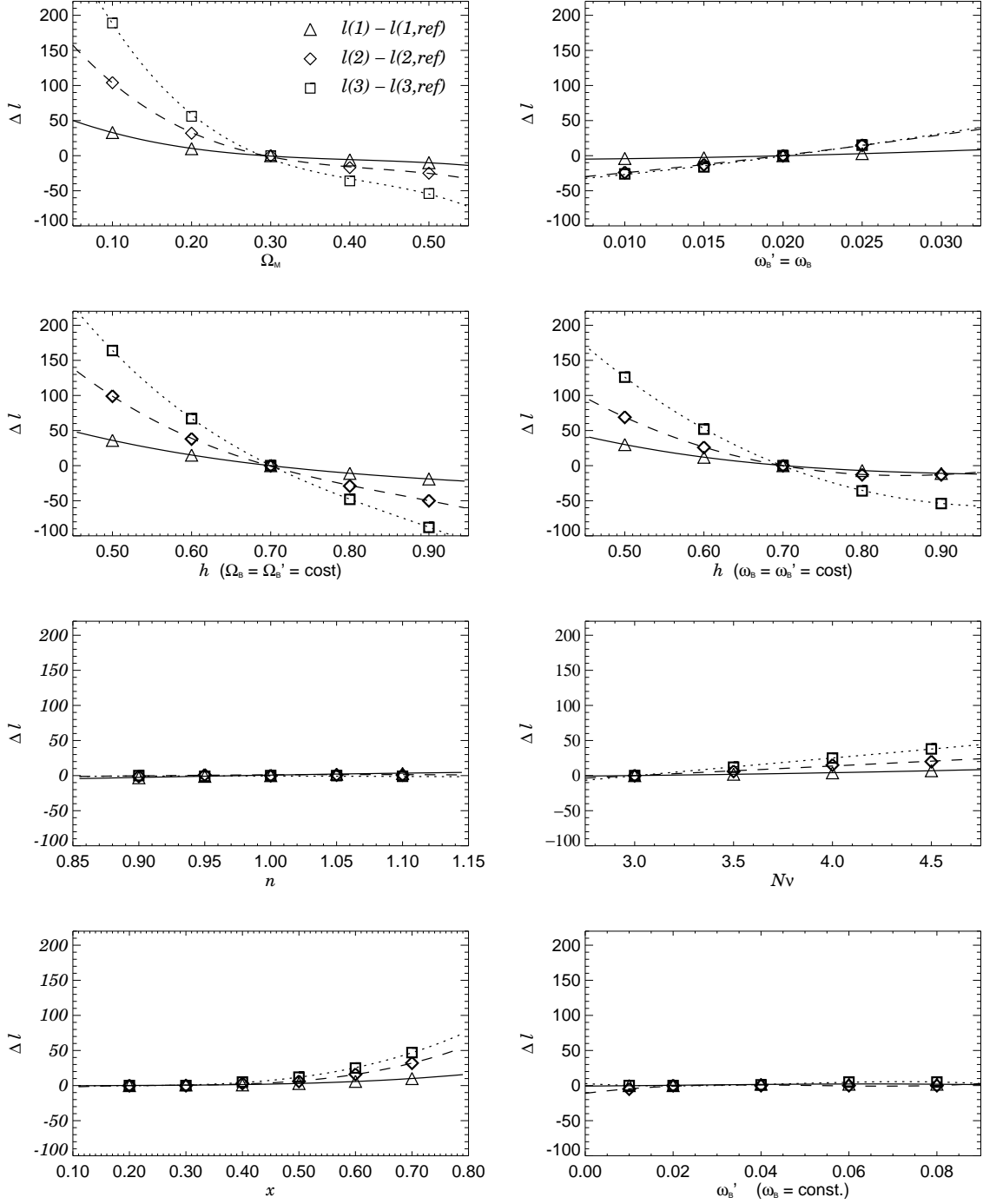}
  \end{center}
\caption{\small Dependences of the locations of the CMB acoustic peaks on 
the values of the cosmological parameters: $ \Omega_m $, 
$ \omega_b = \omega_b' $, $ h $ with 
$ \Omega_b = \Omega_b' = {\rm const.} $, $ h $ with 
$ \omega_b = \omega_b' = {\rm const.} $, $ n $, $ N_\nu $, $ x $, 
and $ \omega_b' $ with $ \omega_b $ constant and $ x = 0.7 $. 
The three indicators used here are the differences of the positions of the first 
three peaks of the models from the ones of the reference model. 
The reference model has: $ \Omega_m = 0.3 $, 
$ \omega_b = \omega_b' = 0.02 $, $ x = 0.2 $, $ h = 0.7 $ and 
$ n = 1.0 $.}
\label{cmblssfig6}
\end{figure}

\begin{figure}[p]
  \begin{center}
    \leavevmode
    \epsfxsize = 15cm
    \epsffile{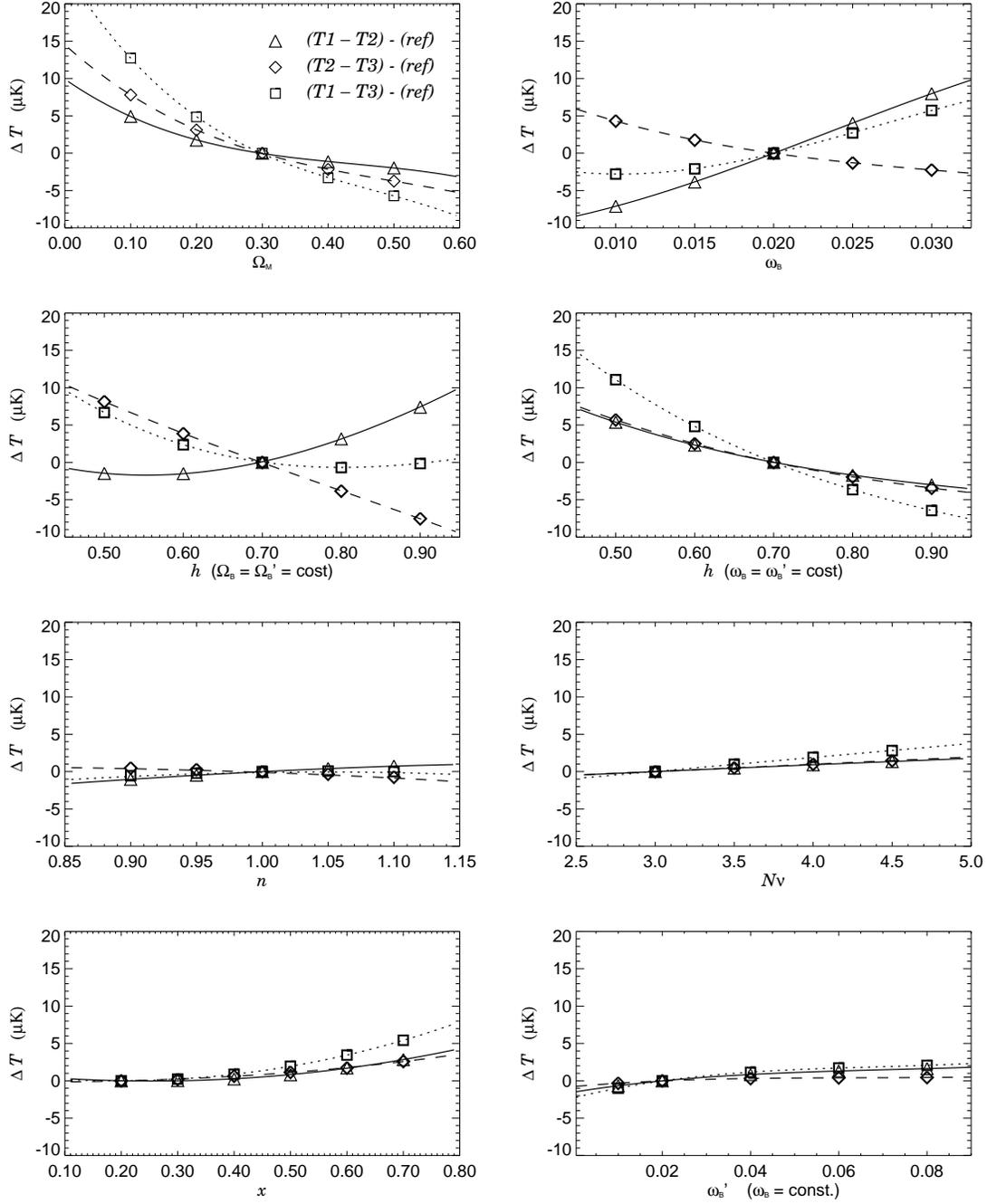}
  \end{center}
\caption{\small Dependences of the temperatures of the CMB acoustic 
peaks on the values of the cosmological parameters: $ \Omega_m $, 
$ \omega_b = \omega_b' $, $ h $ with 
$ \Omega_b = \Omega_b' = {\rm const.} $, $ h $ with 
$ \omega_b = \omega_b' = {\rm const.} $, $ n $, $ N_\nu $, $ x $, 
and $ \omega_b' $ with $ \omega_b $ constant and $ x = 0.7 $. 
The three indicators used here are the deviations of the differences between 
the temperatures of the peaks from the same quantities obtained for a 
reference model. 
The reference model has: $ \Omega_m = 0.3 $, 
$ \omega_b = \omega_b' = 0.02 $, $ x = 0.2 $, $ h = 0.7 $ and 
$ n = 1.0 $.}
\label{cmblssfig7}
\end{figure}

Starting from the reference model of parameters $ \Omega_m = 0.3 $, 
$ \omega_b = \omega_b' = 0.02 $, $ x = 0.2 $, $ h = 0.7 $ and 
$ n = 1.0 $, we change one parameter each time, compute the respective 
models, and plot the CMB and LSS power spectra in order to show the 
dependence on it (figures \ref{cmblssfig8} - \ref{cmblssfig17}). 
Then, we compute the relative locations and heights of the first three 
acoustic peaks of the CMB angular power spectrum and plot them in figures 
\ref{cmblssfig6} and \ref{cmblssfig7} in order to compare their sensitivities 
to the parameters.

In the following we briefly analyze the dependences on every parameter, 
referring to the figure where the respective models are plotted.

\begin{itemize}

\item{\it Matter density} (fig.~\ref{cmblssfig8}): $ \Omega_m $ varies 
from 0.1 to 0.5. 
In flat models a decrease in $ \Omega_m $ implies two things: an 
increase in $ \Omega_\Lambda $ (with the consequent delay in 
matter-radiation equality) and a decrease in $ \Omega_{CDM} $ (if we 
leave unchanged the O and M baryon densities). 
Both these things correspond to boosting and shifting effects on the 
acoustic peaks, while the matter power spectrum goes down, given the 
decreasing of the density of the collisionless species (CDM) and the 
progressive relative growth of the baryon densities, which are responsible 
for the oscillatory features.

\item{\it O baryon density} (fig.~\ref{cmblssfig10}): 
$ \omega_b = \omega_b' $ varies from 0.01 to 0.03. 
An increase of the baryon fraction increases odd peaks (compression phase 
of the baryon-photon fluid) due to extra-gravity from baryons with respect to 
the even peaks (rarefaction phase of the fluid oscillation) in the CMB, and 
generate deeper oscillations in the LSS. 
In particular, the relative magnitudes of the first and second acoustic peaks 
are sensitive to $ \omega_b $, as we see also in figure \ref{cmblssfig7}. 
These effects are completely due to O baryons. 
In fact, even if not shown, we have verified that an increase of 
$ \omega_b $ with a constant $ \omega_b' $ has exactly the 
same consequences for this value of $ x $, while the effect of M baryons 
becomes relevant if we raise the temperature of the mirror sector.

\item {\it Hubble constant} (figs.~\ref{cmblssfig11} and \ref{cmblssfig12}): 
$ h $ varies from 0.50 to 0.90, but now we can leave constant either 
$ \Omega_b = \Omega_b' $ (fig.~\ref{cmblssfig11}) or 
$ \omega_b = \omega_b' $ (fig.~\ref{cmblssfig12}). 
In both cases a decrease in $ h $ corresponds to a delay in the epoch of 
matter-radiation equality and to a different expansion rate. 
This boosts the CMB peaks and slightly changes their location toward 
higher $ \ell $'s (similar to the effect of an increase of $ \Omega_\Lambda $), 
and induces a decrease in the LSS spectrum. 
There are slight differences between the two situations of $ \Omega $ or 
$ \omega $ constant, evident in particular on the first acoustic peak and on 
the matter oscillations. 
If fact, when we consider $ \omega_{b,\:b'} $ constant, the baryon 
densities $ \Omega_{b,\:b'} = \omega_{b,\:b'} / h^2 $ grow for 
decreasing $ h $, then favouring the raise of the first peak in the CMB and 
the onset of oscillations in the LSS.

\item {\it Spectral index} (fig.~\ref{cmblssfig13}): $ n $ varies from 0.90 to 
1.10. Increasing $ n $ will raise the power spectra at large $ \ell $'s with 
respect to the low $ \ell $'s and at large values of $ k $ with respect to low 
values. 
This is not evident in figure (except before the first acoustic peak), where 
the curves seem nearly parallel as if they were simply vertically shifted; 
this means a low sensitivity to the spectral index in this range, as also 
evident in figures \ref{cmblssfig6} and \ref{cmblssfig7} for the CMB.

\item {\it Extra-neutrino species} (fig.~\ref{cmblssfig14}): $ \Delta N_\nu $ 
varies from 0.0 to 1.5. 
The effect of increasing the number of massless neutrino species is a slow 
raise of the first acoustic peak and a shift to higher $ \ell $ values for next 
peaks, together with a slight lowering of the matter power spectrum; all 
these changes are nearly proportional to $ \Delta N_\nu $, as shown also in 
figures \ref{cmblssfig6} and \ref{cmblssfig7}.

\item {\it Ratio of temperatures} (fig.~\ref{cmblssfig15}): $ x $ varies from 0.2 
to 0.7. 
Concerning the CMB, the effect of raising $ x $ is qualitatively the same as 
an increase in $ \Delta N_\nu $, but more pronounced (for these 
``cosmologically compatible'' ranges) and with a non-linear dependence. 
In the LSS spectrum, instead, the situation is different from the case of 
extra-neutrino species, as now a growth of $ x $ induces the onset of the 
oscillatory features at lower values of $ k $. 
These behaviours have been studied in more details in \S~\ref{cmb_2} and 
\S~\ref{lss_2}.

\item {\it M baryon density} (fig.~\ref{cmblssfig17}): $ \omega_b' $ 
varies from 0.01 ($ \omega_b $ / 2) to 0.08 ($ 4 \omega_b $). 
The value of $ x $ is now raised to 0.7, because for 0.2 there aren't 
differences between models with different $ \omega_b' $ values
(we start observing small deviations only for the higher 
$ k $-values in the matter power spectrum). 
Also this behaviour has been studied in more details in \S~\ref{cmb_2} and 
\S~\ref{lss_2}; here we want to emphasize the low sensibility of the CMB on 
$ \omega_b' $ (with a slightly stronger dependence starting from the 
third peak) and, on the contrary, the high sensitivity of the LSS. 
For the first one, an increase in $ \omega_b' $ causes a very low 
increase of the height of the first peak and a progressive more pronounced 
decrease for the next peaks, while for the second one there is a rapid 
deepening of the oscillations, slightly changing their locations.

\end{itemize}

In figures \ref{cmblssfig6} and \ref{cmblssfig7} we focus our attention on the 
CMB first three peaks, choosing some indicator which could quantify the 
sensitivity to the parameters previously discussed. 
In figure \ref{cmblssfig6} we analyze the locations of the peaks, plotting the 
differences of the locations between the various models and the reference 
one for the three peaks; in figure \ref{cmblssfig7} we plot the deviations 
of the differences between the heights of the peaks from the same quantities 
obtained for the reference model. 
In this way we obtain a clear picture of the trends of these indicators varying 
the parameters. 
These plots provide a useful reference in order to evaluate the influence of 
each parameter on the CMB and LSS power spectra, and they contain a 
number of informations; we can extract some of them particularly worth of 
noting.

Looking at the locations, we see a great sensitivity on the matter density 
and the Hubble constant, and a negligible one on the spectral index and the 
amount of mirror baryons. 
Concerning the extra-neutrino species and the temperature of the sectors, 
the sensitivities are comparable, but the trends are different: they are 
respectively a constant slope for $ \Delta N_\nu $ and an increasing one 
for $ x $.

As regards the peak temperatures, the most sensitive parameter, besides 
$ \Omega_m $ and $ h $, is $ \omega_b $; the dependence on 
$ n $ and $ \omega_b' $ is a bit greater than what it is for the 
locations, and the differences between the trends with $ N_\nu $ and $ x $ 
are slightly more evident, specially for values $ x > 0.6 $.

\newpage
\section{Comparison with observations}
\label{comp_obs}

So far we have studied the behaviour of the photon and matter power 
spectra varying many cosmological parameters with special attention to 
the two mirror parameters, i.e. the ratio of the temperatures of two sectors 
$ x $ and the amount of mirror baryons $ \omega_b' $.

Here we want to compare these models with some experimental data, in 
order to estimate the compatibility of the mirror scenario with observations 
and possibly reduce the parameter ranges.

As written in \S~\ref{mirror_mod}, we are not able to fit the parameters now. 
It is due to the slowness of our present version of the numerical code, but 
our game will be to choose some representative model and compare it with 
observations.

In the last decade the anisotropies observed in the CMB temperature 
became the most important source of information on the cosmological 
parameters: a lot of experiments (ground-based, balloon and satellite) were 
dedicated to its measurement. 
At the same time, many authors proved that its joint analysis with the 
fluctuations in the matter distribution (they have both the same primordial 
origin) are a powerful instrument to determine the parameters of the 
Universe. 
As in \S~\ref{cmb_2} and \S~\ref{lss_2}, we analyze separately the variation 
of $ x $ and $ \omega_b' $ in the mirror models, using now both the 
CMB and LSS informations at the same time.

In order to compare our predictions with observations, we use data among 
the best ones available at present: for the CMB the WMAP \cite{wmap-data} 
and ACBAR \cite{acbar-data} data, and for the LSS the 2dF survey (in 
particular the binned power spectrum obtained by Tegmark et al. 
\cite{2df-teg}).
In order to compare with the standard CDM results, we choose a reference 
cosmological model with scalar adiabatic perturbations and no massive 
neutrinos with the following set of parameters \cite{wmap-par}:
$ \Omega_m = 0.25, ~~ \omega_b = 0.023, ~~ 
\Omega_\Lambda = 0.75, ~~ h = 0.73, ~~ n = 0.97 $.
As usually, we include in this model the mirror sector; 
for the sake of comparison, in all calculations the total amount of matter 
$ \Omega_m = \Omega_{CDM} + \Omega_b + \Omega'_b$
is maintained constant. 
Mirror baryons contribution is thus always increased at the 
expenses of diminishing the CDM contribution.

\begin{figure}[p]
  \begin{center}
    \leavevmode
    \epsfxsize = 13cm
    \epsffile{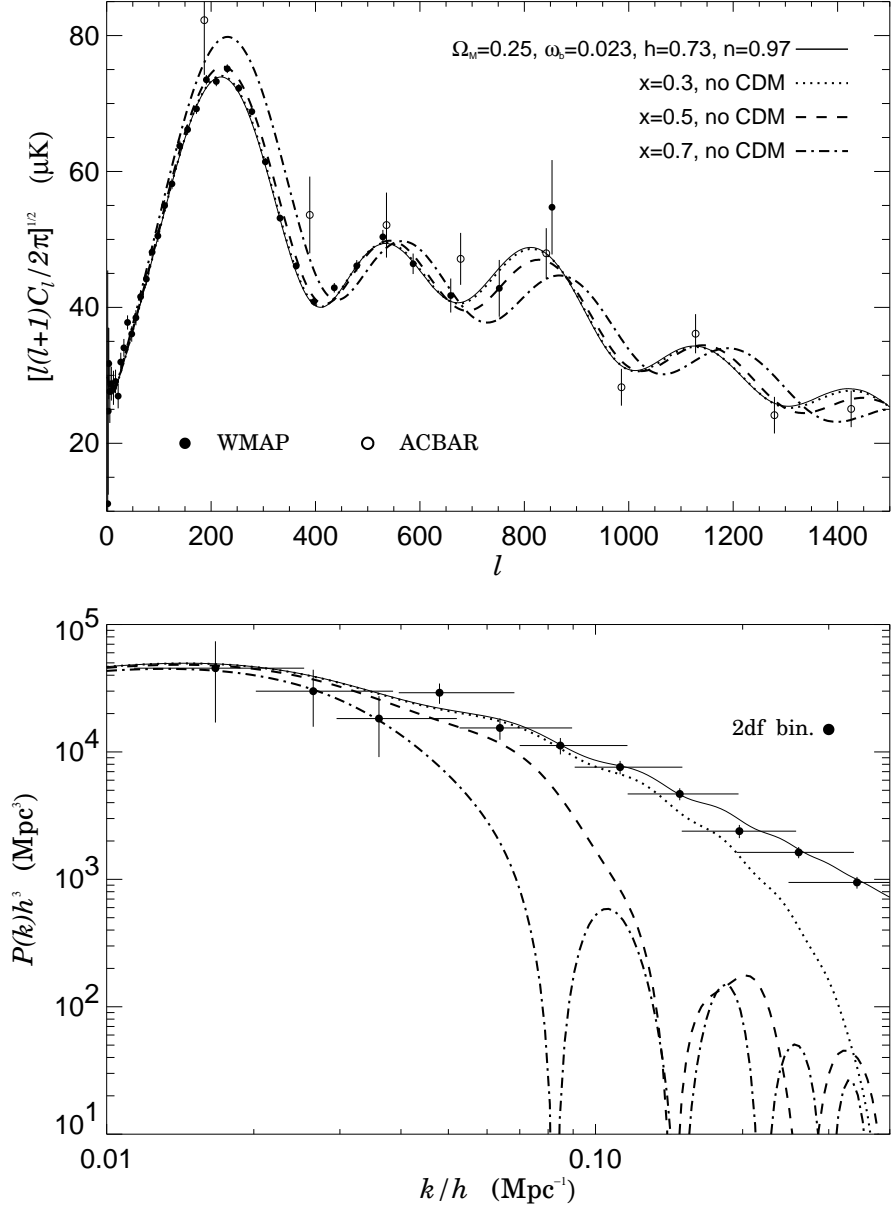}
  \end{center}
\addvspace{1.2cm}
\caption{\small CMB and LSS power spectra for various mirror models with 
different values of $ x $, compared with observations and with 
a standard reference model (solid line) of 
parameters $ \Omega_0 = 1 $, $ \Omega_m = 0.25 $, 
$ \Omega_\Lambda = 0.75 $, 
$ \omega_b = \Omega_b h^2 = 0.023 $, $ h = 0.73 $, 
$ n = 0.97 $. 
The mirror models have the same parameters as the standard one, but with 
$ x = 0.3, 0.5, 0.7 $ and 
$ \omega_b' = \Omega_m h^2 - \omega_b $ (no CDM) for all 
models. 
{\sl Top panel.} Comparison of the photon power spectrum with the WMAP 
and ACBAR data. 
{\sl Bottom panel.} Comparison of the matter power spectrum with the 2dF 
binned data.}
\label{cmblssfig18}
\end{figure}

We start from figure \ref{cmblssfig18}, where 
we assume that the dark matter is entirely due to mirror baryons and 
we consider variations of the $ x $ parameter
(as in the upper part of figures \ref{cmblssfig1} and \ref{cmblssfig3}). 
In top panel, we see that with the accuracy of the current anisotropy 
measurements the CMB power spectra for mirror models are perfectly 
compatible with data, except for the higher-$ x $ one. 
Indeed, the deviations from the standard model are weak for $ x \lsim 0.5 $, 
even in a Universe full of mirror baryons (see \S~\ref{cmb_2}). 
In lower panel, instead, the situation is very different: oscillations due 
to mirror baryons are too deep to be in agreement with data, and only 
models with low values of $ x $ (namely $ x \lsim 0.3 $) are acceptable. 
Thus, we find the first strong constraint on 
the mirror parameter space: models with high mirror sector temperatures 
and all the dark matter made of mirror baryons have to be excluded.

\begin{figure}[p]
  \begin{center}
    \leavevmode
    \epsfxsize = 13cm
    \epsffile{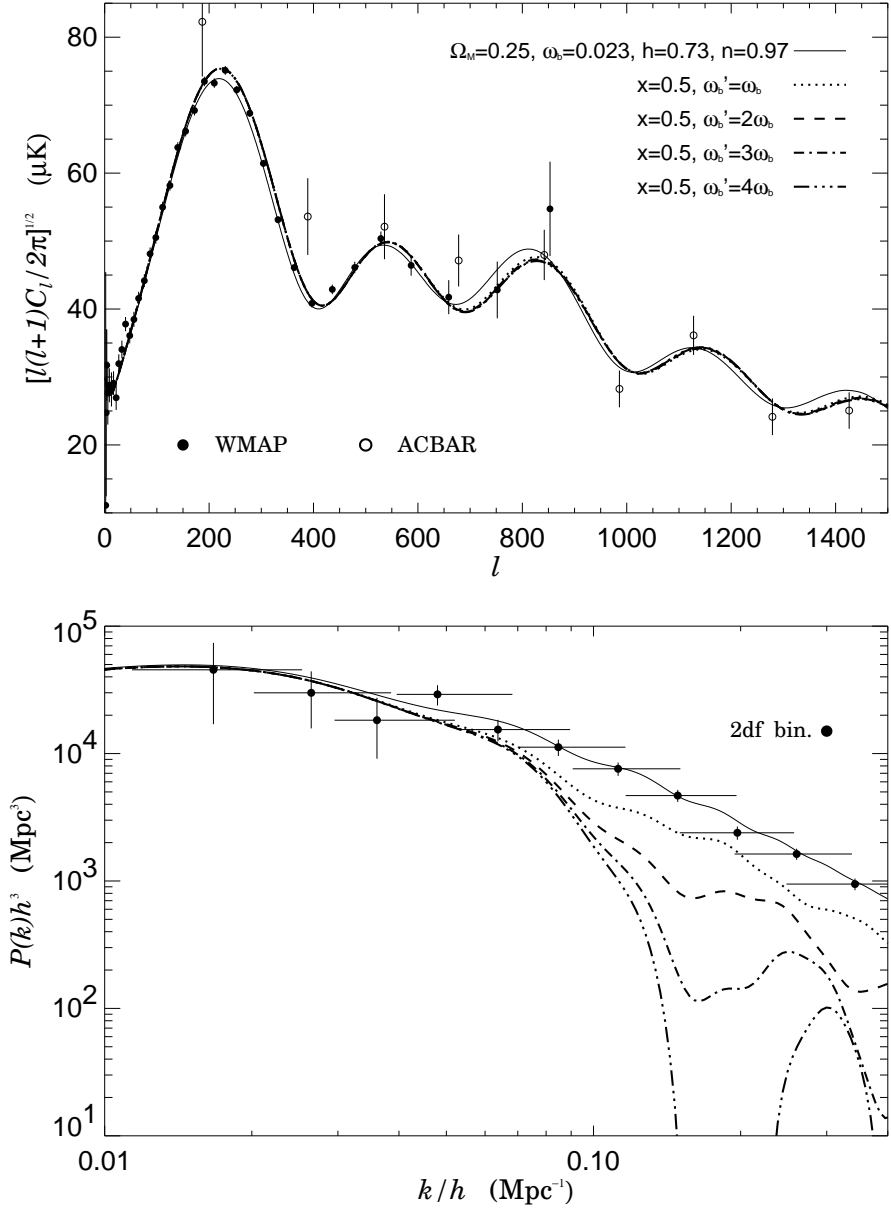}
  \end{center}
\addvspace{1.2cm}
\caption{\small CMB and LSS power spectra for various mirror models with 
different values of mirror baryon density, compared with observations and 
with a standard reference model (solid line) of 
parameters $ \Omega_0 = 1 $, $ \Omega_m = 0.25 $, 
$ \Omega_\Lambda = 0.75 $, 
$ \omega_b = \Omega_b h^2 = 0.023 $, $ h = 0.73 $, 
$ n = 0.97 $. 
The mirror models have the same parameters as the standard one, but 
with $ x = 0.5 $ and for 
$ \omega_b' = \omega_b, 2 \omega_b, 3 \omega_b, 4 \omega_b $. 
{\sl Top panel.} Comparison of the photon power spectrum with the WMAP 
and ACBAR data. 
{\sl Bottom panel.} Comparison of the matter power spectrum with the 
2dF binned data.}
\label{cmblssfig19}
\end{figure}

In figure \ref{cmblssfig19} we compare with observations models 
with the same $ x $, but different mirror baryon contents 
(as in the bottom panels of figures \ref{cmblssfig1} and \ref{cmblssfig3}). 
The above mentioned low sensitivity of the CMB power spectra on 
$ \omega_b' $ doesn't give us indications for this parameter (even for 
high values of $ x $), but the LSS power spectrum helps us again, 
confirming a sensitivity to the mirror parameters greater than the CMB one. 
This is another example of the great advantage of a joint analysis of CMB 
and LSS power spectra, being the following conclusion impossible looking 
at the CMB only. 
This plot tells us that also high values of $ x $ can be compatible with 
observations if we decrease the amount of mirror baryons in the Universe. 
It is a second useful indication: in case of high mirror sector temperatures 
we have to change the mirror baryon density in order to reproduce 
the oscillations present in the LSS data.

Therefore, after the comparison with experimental data, we are left with 
three possibilities for the Mirror Universe parameters:
\begin{itemize}
\item high $ x \lsim 0.5 $ and low $ \omega_b' $ (differences from the CDM in 
the CMB, and oscillations in the LSS with a depth modulated by the baryon 
density);
\item low $ x \lsim 0.3 $ and high $ \omega_b' $ (completely equivalent to the 
CDM for the CMB, and few differences for the LSS in the linear region);
\item low $ x $ and low $ \omega_b' $ (completely equivalent to the 
CDM for the CMB, and nearly equivalent for the LSS in the linear region and 
beyond, according to the mirror baryon density).
\end{itemize}
Thus, with the current experimental accuracy, we can exclude only models 
with high $ x $ and high $ \omega_b' $.
Our next step will be to consider some interesting mirror models and 
compute their power spectra. 

\begin{figure}[p]
  \begin{center}
    \leavevmode
    \epsfxsize = 13cm
    \epsffile{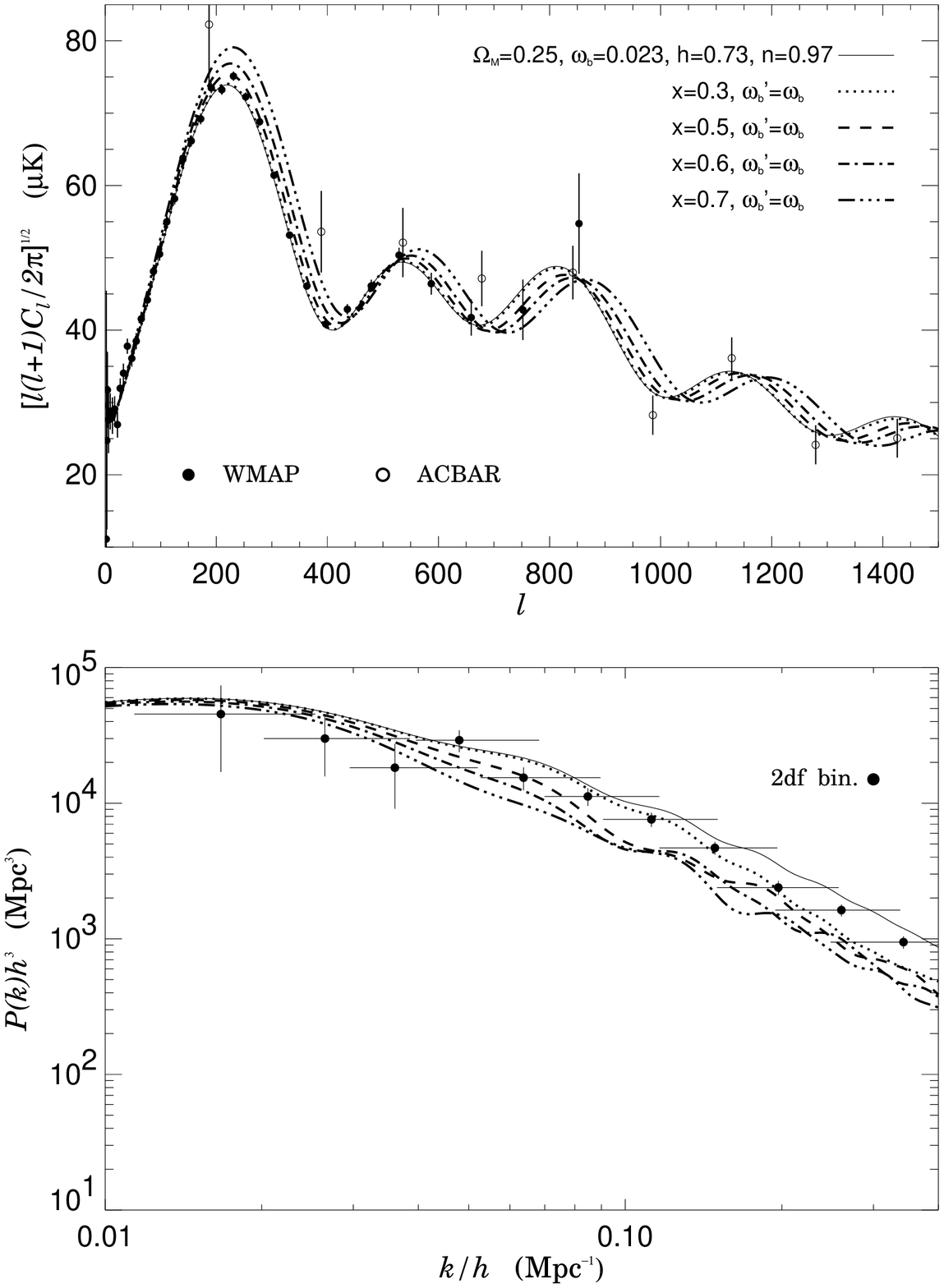}
  \end{center}
\addvspace{1.2cm}
\caption{\small CMB and LSS power spectra for various mirror models with 
different values of $ x $ and equal amounts of ordinary and mirror baryons, 
compared with observations and with 
a standard reference model (solid line) of 
parameters $ \Omega_0 = 1 $, $ \Omega_m = 0.25 $, 
$ \Omega_\Lambda = 0.75 $, 
$ \omega_b = \Omega_b h^2 = 0.023 $, $ h = 0.73 $, 
$ n = 0.97 $. 
The mirror models have the same parameters as the standard one, but with 
$ \omega_b' = \omega_b $ and $ x = 0.3, 0.5, 0.6, 0.7 $. 
{\sl Top panel.} Comparison of the photon power spectrum with the WMAP 
and ACBAR data. 
{\sl Bottom panel.} Comparison of the matter power spectrum with the 
2dF binned data.}
\label{cmblssfig20}
\end{figure}

In figure \ref{cmblssfig20} we plot models with equal amounts of ordinary 
and mirror baryons and a large range of temperatures. 
This is an interesting situation, because the case 
$ \Omega_b' = \Omega_b $ could be favoured in some baryogenesis 
scenario with some mechanism that naturally lead to equal baryon number 
densities in both visible and hidden sectors \cite{baryo-lepto}. 
These models are even more interesting when we consider both their CMB 
and LSS power spectra. 
In top panel of figure we see that the temperature anisotropy spectra are 
fully compatible with observations until $ x \simeq 0.5 $, 
without large deviations from the standard case. 
In bottom panel, we have a similar situation for the matter power spectra, 
with some oscillations and a slightly greater slope, that could be useful to 
better fit the oscillations present in the data and to solve the discussed 
problem of the desired cutoff at low scales. 
Let us observe that we are deliberately neglecting the biasing problem, 
given that an indication on its value can come only from a fit of the 
parameters; then, we have indeed a small freedom to vertically shift the 
curves in order to better fit the experimental data.

\begin{figure}[p]
  \begin{center}
    \leavevmode
    \epsfxsize = 13cm
    \epsffile{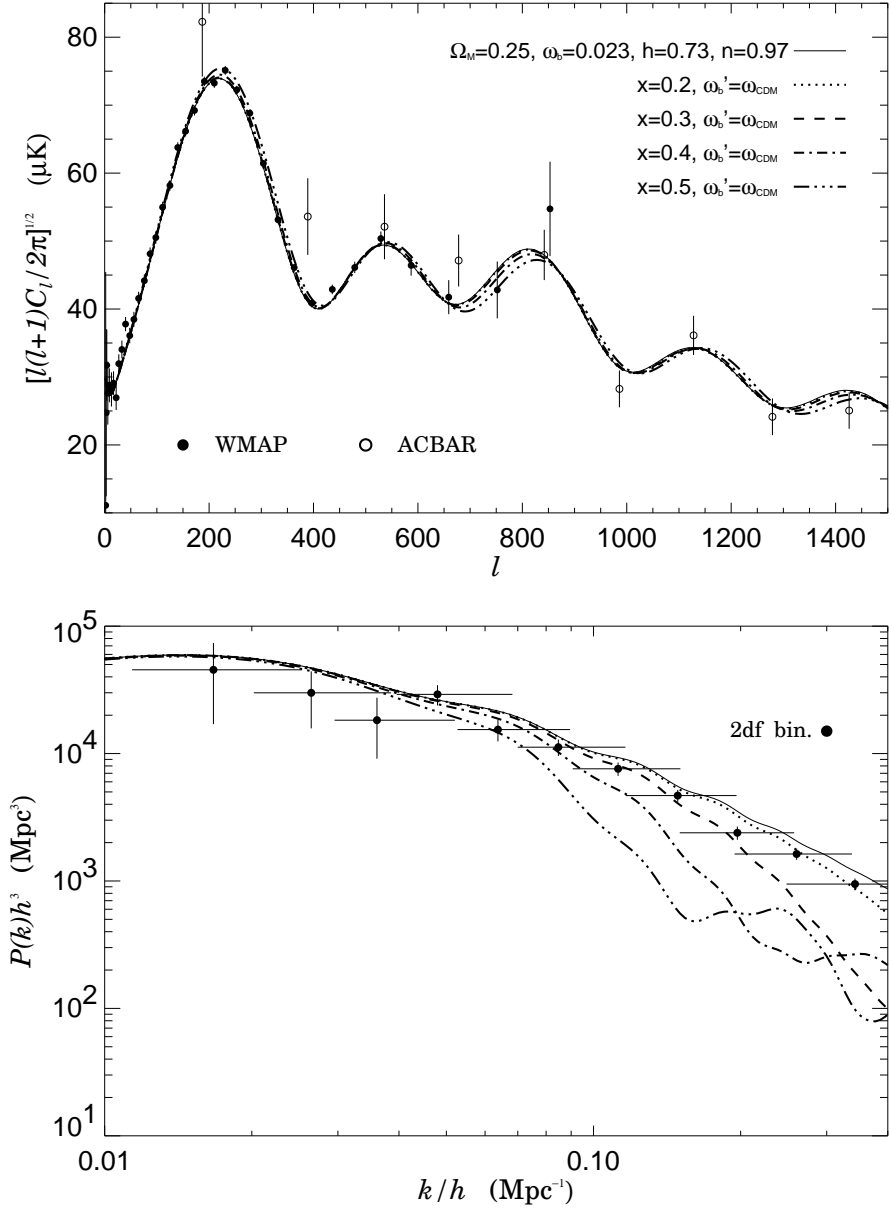}
  \end{center}
\addvspace{1.2cm}
\caption{\small CMB and LSS power spectra for various mirror models with 
different values of $ x $ and equal amounts of CDM and mirror baryons, 
compared with observations and with 
a standard reference model (solid line) of 
parameters $ \Omega_0 = 1 $, $ \Omega_m = 0.25 $, 
$ \Omega_\Lambda = 0.75 $, 
$ \omega_b = \Omega_b h^2 = 0.023 $, $ h = 0.73 $, 
$ n = 0.97 $. 
The mirror models have the same parameters as the standard one, but with 
$ \omega_b' = \omega_{CDM} $ and $ x = 0.2, 0.3, 0.4, 0.5 $. 
{\sl Top panel.} Comparison of the photon power spectrum with the WMAP 
and ACBAR data. 
{\sl Bottom panel.} Comparison of the matter power spectrum with the 
2dF binned data.}
\label{cmblssfig21}
\end{figure}

Models of a Mirror Universe where the dark matter is composed in equal 
parts by CDM and mirror baryons are plotted in figure \ref{cmblssfig21}. 
Here we concentrate on $ x $-values lower than the previous figure, because 
the greater mirror baryonic density would generate too many oscillations in 
the linear region of the matter power spectrum. 
In top panel we show that, apart from little deviations for the model with 
higher $ x $, all other models are practically the same. 
In bottom panel, instead, deviations are big, and we can still use LSS as a 
test for models. 
Indeed, models with $ x \gsim 0.4 $ are probably to exclude, even taking 
into account a possible bias. 
Models with lower $ x $ are all consistent with observations.

Observing the figures, even if it is surely premature (given the lack of a 
detailed statistical analysis of mirror models), we are tempted to guess that 
mirror baryons could hopefully better 
reproduce the oscillations present in the LSS power spectra. 
Thus, we are waiting for more accurate data (in particular for the large scale 
structure) in order to obtain indications for a standard or a Mirror Universe.

\newpage
\section{Conclusions}

We have discussed cosmological implications, in terms of cosmic 
microwave background radiation and large scale structure, of the 
parallel mirror world with the same microphysics as the ordinary one, 
but having smaller temperature, $ T' < T $, with the limit $ T' / T < 0.64 $ 
set by the BBN constraints. 
This bound implies that the relativistic contribution to the cosmological 
energy density is dominated by the ordinary sector, while for the matter 
component the complementary situation can occur and the dark matter of 
the Universe can be made, in part or entirely, of mirror baryonic dark 
matter (MBDM). 

In the first paper of this series \cite{paper1} we showed that, since the 
existence of a mirror hidden sector changes the time of key epochs and the 
nature of dark matter, there are important consequences in the structure 
formation scenario for a Mirror Universe. 
We studied the possible mirror scenarios in presence of adiabatic scalar 
density perturbations in the context of the Jeans gravitational instability 
theory, finding important differences from the cold dark matter model.

Here, in order to evaluate the effects of a mirror sector on the CMB and LSS, 
we modified a numerical code existing for the standard Universe 
in order to take into account a hidden mirror sector; 
in this way we were able to predict the expected power spectra of cosmic 
microwave background and large scale structure for a flat Mirror Universe 
with adiabatic scalar density perturbations. 
In this MBDM scenario we studied the dependence of power spectra on 
mirror parameters $ x $ 
and $ \Omega_b' $, and also on other cosmological parameters, as 
$ \Omega_m $, $ \Omega_b $, $ h $, $ n $, $ N_\nu $.
We enlarged our previous studies on this topic \cite{paolo}, 
considering a much larger set of models.
This allowed us to study in detail 
the sensitivity of mirror CMB and LSS spectra on the cosmological 
parameters and to show some interesting mirror model which could be 
a viable alternative to the standard ones.

In CMB spectra we found various differences from a so-called 
standard (CDM) reference model for $ x \gsim 0.3 $ and a non-linear 
dependence on $ x $, specially evident in the first and third peaks. 
The dependence on the mirror baryon density is instead very low. 
We computed also the power spectrum of mirror CMB photons, even if, 
unfortunately, by definition we can't 
reveal them and we cannot exchange the information with the mirror 
physicists and combine our observations. 

Turning to the LSS power spectra, we showed an influence of the mirror 
sector bigger than for the CMB, with a great dependence on both mirror 
temperature and baryonic density. 
Both of them control the oscillations generated in power spectrum, but the 
first one 
influences the scale at which they start, while the second one their depth. 
In this case the mirror sector effects are evident also for low $ x $-values, 
if we don't take a too small value for $ \Omega_b' $.
We extended the models also to smaller (non linear) scales, in order to 
show the cutoff present in the mirror scenario, due to the existence of 
the dissipative mirror Silk scale. 
We demonstrated the existence of this cutoff, specially dependent on 
$ x $-value, but modulated also by $ \Omega_b' $. 
This is an important feature of the mirror structure formation scenario, 
because it could explain the observed small number of substructures 
which is a problem for cold dark matter.
Furthermore, we showed that for low $ x $-values, $ x \lsim 0.3 $, MBDM is 
equivalent to CDM for the CMB and LSS at linear scales.
This is an interesting opportunity for mirror matter, since for low mirror 
temperatures we could obtain models completely equivalent to the CDM 
scenario at larger scales (when it works well), but with less power at smaller 
scales (when it has problems).

In addition, we have shown the dependence of the photon and matter 
power spectra on the used cosmological parameters, in order to know 
the relative sensitivities. 
For the CMB case, we used also some indicators, as the differences 
between the heights and positions of the peaks compared 
with the same quantities computed for a reference model.

Our predictions have been compared with the observations 
(the WMAP \cite{wmap-data} and ACBAR \cite{acbar-data} data for the CMB, 
and the 2dF binned data \cite{2df-teg} for the LSS)
in order to obtain bounds on the possible existence of 
the mirror sector. 
In this stage we jointly considered both CMB and LSS data, 
obtaining important limits on the mirror parameter space. 
In fact, just from a visual inspection we easily found that mirror models with 
high $ x $ and high $ \Omega_b' $ are excluded by LSS observations, 
because they generate too deep oscillations in power spectra. 

Even if a statistical analysis is necessary in order to extract detailed 
informations from the experimental data, nevertheless we can reach some 
general conclusion in a rather straightforward way:
 
\begin{itemize}

\item The present LSS data are not compatible with a scenario where all the 
dark matter is made of mirror baryons, unless we consider enough small 
values of $ x $: 
$ x \lsim 0.3 \sim x_{\rm eq} $.

\item High values of $ x $, $ x > 0.5 $, can be excluded
even for a relatively small amount of mirror baryons. 
In fact, we observe relevant effects on LSS and CMB power spectra down to 
values of M baryon density of the order $ \Omega'_b \sim \Omega_b $. 

\item Intermediate values of $ x $, $ 0.3 < x < 0.5 $, can be 
allowed if the MBDM is a subdominant component of dark matter, 
$ \Omega'_b \lsim \Omega_b \lsim \Omega_{CDM} $. 

\item For small values of $ x $, $ x < 0.3 $, 
the MBDM and the CDM scenarios are indistinguishable as concerns 
the CMB and the linear LSS power spectra.
In this case, in fact, the mirror Jeans and Silk lengths, 
which mark region of the spectrum where the effects of 
mirror baryons are visible, decrease to very low values, which undergo 
non linear growth from relatively large redshift (for details see Paper 1 
\cite{paper1}). 

\end{itemize}

Thus, with the current experimental accuracy, we can exclude only 
models with high $ x $ and high $ \Omega_b' $; 
however, there can be many possibilities to disentangle the cosmological 
scenario of two parallel worlds with the future high precision data 
concerning the large scale structure, CMB anisotropy, structure of the 
galaxy halos, gravitational microlensing, oscillation of 
observable particles into their mirror partners, etc.  

The aim of these series of papers was contained in a question: 
{\sl ``is mirror matter still a reliable dark matter candidate?''}
Now, at the end of this second paper, we reached a partial answer: 
{\sl on the light of current observations of CMB and LSS at linear 
scales the mirror baryonic dark matter is fully in agreement with 
experimental data}. 
In addition, {\sl we obtained some useful constraints on the mirror 
parameter space}, that can address our future efforts to understand 
other aspects of the Mirror Universe. 
Furthermore, so far the mirror dark matter candidate still shows 
interesting potentialities to solve some open problems of the ``standard'' 
cosmological scenario.


\section*{Acknowledgements}

\noindent 
I am grateful to my invaluable collaborators 
Zurab Berezhiani, Denis Comelli and Francesco Villante. 
I would like to thank also 
Silvio Bonometto, Stefano Borgani, Alfonso 
Cavaliere and Nicola Vittorio for interesting discussions.




\begin{thebibliography}{2}

\bibitem{paper1}
P. Ciarcelluti, astro-ph/0409630.

\bibitem{broken}
Z. Berezhiani, Acta Phys. Polon. B 27 (1996) 1503; \\
R. Foot, H. Lew and R. R. Volkas, JHEP 007, 032 (2000) 
[hep-ph/0006027]; \\
R. N. Mohapatra, S. Nussinov and V. L. Teplitz,
Phys. Rev. D 66, 063002 (2002) [hep-ph/0111381].

\bibitem{mirror}
R. Foot, H Lew and R. R. Volkas, Phys. Lett. B 272, 67 (1991). \\
The idea of mirror particles was earlier discussed in: \\
T. D. Lee and C. N. Yang, Phys. Rev. 104, 256 (1956); \\
I. Kobzarev, L. Okun and I. Pomeranchuk, Sov. J. Nucl. Phys. 3, 837 (1966); \\
M. Pavsic, Int. J. Theor. Phys. 9, 229 (1974). \\
For a review: \\
R. Foot, hep-ph/0207175; \\
R. Foot, {\it Shadowlands, quest for mirror matter in the Universe}, 
Universal Publishers, Parkland FL, 2002; \\
R. Foot, astro-ph/0407623; \\
Z. Berezhiani, Int. J. Mod. Phys. A 19, 3775 (2004) [hep-ph/0312335].

\bibitem{blinkhlo}
S. I. Blinnikov and M. Yu. Khlopov, Sov. J. Nucl. Phys. 36, 472 (1982); \\
S. I. Blinnikov and M. Yu. Khlopov, Sov. Astron. 27, 371 (1983).

\bibitem{mixing}
B. Holdom, Phys. Lett. B 166, 196 (1986); \\
E. Carlson and S. Glashow, Phys. Lett. B193, 168 (1987); \\
R. Foot and X-G. He, Phys. Lett. B 267, 509 (1991); \\
M. Collie and R. Foot, Phys. Lett. B 432, 134 (1998) [hep-ph/9803261]; \\
A. Yu. Ignatiev and R. R. Volkas, Phys. Lett. B 487, 294 (2000)
[hep-ph/0005238];\\
R. Foot, A. Yu. Ignatiev and R. R. Volkas, Phys. Lett. B 503, 355 (2001)
[astro-ph/0011156].

\bibitem{neutrino} 
E. Akhmedov, Z. Berezhiani and G. Senjanovi\'c,
Phys. Rev. Lett. 69, 3013 (1992); \\
R. Foot, Mod. Phys. Lett. A 9, 169 (1994) [hep-ph/9402241];\\
R. Foot and R. R. Volkas, Phys. Rev. D 52, 6595 (1995) [hep-ph/9505359]; \\
Z. Berezhiani, R.N. Mohapatra, Phys. Rev. D 52, 6607 (1995)
[hep-ph/9505385]; \\
Z. Silagadze, Phys. Atom. Nucl. 60, 272 (1997) [hep-ph/9503481]; \\
R. Foot, R.R. Volkas, Phys. Rev. D 61, 043507 (2000) [hep-ph/9904336]; \\
V. Berezinsky, M. Narayan, F. Vissani, Nucl. Phys. B 658, 254 (2003)
[hep-ph/0210204].

\bibitem{Macho}
S. Blinnikov, astro-ph/9801015; \\
R. Foot, Phys. Lett. B 452, 83 (1999) [astro-ph/9902065]; \\
R.N. Mohapatra, V. Teplitz, Phys. Lett. B 462, 302 (1999) [astro-ph/9902085]. 

\bibitem{mir_GRB}
S. Blinnikov, astro-ph/9902305; \\
R. Volkas, Y. Wong, Astropart. Phys. 13, 21 (2000) [astro-ph/9907161]; \\
R.N. Mohapatra, S. Nussinov and V.L. Teplitz, 
Astropart. Phys. 13, 295 (2000) [astro-ph/9909376]; \\
S. Blinnikov, Surveys High Energ. Phys. 15, 37 (2000) [astro-ph/9911138]; \\
R. Foot and Z. K. Silagadze, astro-ph/0404515.

\bibitem{assione}
Z. Berezhiani, Phys. Lett. B 417, 287 (1998) [hep-ph/9609342]; \\ 
Z. Berezhiani, L. Gianfagna and M. Giannotti, Phys. Lett. B 500, 286 (2001) 
[hep-ph/0009290]; \\ 
L. Gianfagna, M. Giannotti and F. Nesti, hep-ph/0409185.

\bibitem{ignavol-lss}
A. Yu. Ignatiev and R. R. Volkas, Phys. Rev. D 68, 023518 (2003)
[hep-ph/0304260].

\bibitem{paolo}
Z. Berezhiani, P. Ciarcelluti, D. Comelli and F. L. Villante, astro-ph/0312605;\\
P. Ciarcelluti, astro-ph/0312607; \\
P. Ciarcelluti, astro-ph/0409629.

\bibitem{mir_halo}
R.N. Mohapatra, V. Teplitz, Astrophys. J. 478, 29 (1997) [astro-ph/9603049]; \\
R.N. Mohapatra, V. Teplitz, Phys. Rev. D 62, 063506 (2000) 
[astro-ph/0001362]; \\
R. Foot and R. R. Volkas, astro-ph/0407522.

\bibitem{ortho}
S. L. Glashow, Phys. Lett. B 167, 35 (1986);\\
R. Foot and S. N. Gninenko, Phys. Lett. B 480, 171 (2000) [hep-ph/0003278]; \\
R. Foot, astro-ph/0309330; \\
S.N. Gninenko, Phys. Lett. B 326 (1994) 317. 

\bibitem{bader}
A. Badertscher {\it et al.}, hep-ex/0311031.

\bibitem{mir_dama}
R. Foot, Phys. Rev. D 69, 036001 (2004) [hep-ph/0308254];\\
R. Foot, astro-ph/0403043; \\
R. Foot, Mod. Phys. Lett.  A 19, 1841 (2004) [astro-ph/0405362].

\bibitem{mir_meteor}
A. Yu. Ignatiev and R. R. Volkas, Phys. Rev. D 62, 023508 (2000)
[hep-ph/0005125]; \\
Z. K. Silagadze, Acta Phys. Pol. B 32, 99 (2001) [hep-ph/0002255]; \\
R. Foot, Acta Phys. Polon. B 32, 3133 (2001) [hep-ph/0107132]; \\
R. Foot and T. L. Yoon, Acta Phys. Polon. 33, 1979 (2002) 
[astro-ph/0203152]; \\
R. Foot and S. Mitra, Astropart. Phys. 19, 739 (2003) [astro-ph/0211067]; \\
A. Yu. Ignatiev and R. R. Volkas, hep-ph/0306120; \\
R. Foot and S. Mitra, Phys. Rev. D 68, 071901 (2003) [hep-ph/0306228]; \\
Z. K. Silagadze, astro-ph/0311337.

\bibitem{detect_mir_frag}
S. Mitra and R. Foot, Phys. Lett. B 558, 9 (2003) [astro-ph/0301229]; \\
R. Foot and S. Mitra, Phys. Lett. A 315, 178 (2003) [cond-mat/0306561].

\bibitem{pioneer}
R. Foot and R. R. Volkas, Phys. Lett. B 517, 13 (2001) [hep-ph/0108051].

\bibitem{mir_planet}
R. Foot, Phys. Lett. B 471, 191 (1999) [astro-ph/9908276]; \\
R. Foot, Phys. Lett. B 505, 1 (2001) [astro-ph/0101055]; \\
R. Foot and Z. K. Silagadze, Acta Phys. Pol. B 32, 2271 (2001)
[astro-ph/0104251]; \\
R. Foot, A. Yu. Ignatiev and R. R. Volkas, Astropart. Phys.
17, 195 (2002) [astro-ph/0010502]; \\
Z. K. Silagadze, Acta Phys. Pol. B 33, 1325 (2002) [astro-ph/0110161]; \\
R. Foot, astro-ph/0406257.

\bibitem{bcv}
Z. Berezhiani, D. Comelli and F. L. Villante, Phys. Lett. B 503, 362 (2001) 
[hep-ph/0008105]. 

\bibitem{inflation}
E. W. Kolb, D. Seckel and M. S. Turner, Nature 314, 415 (1985);\\
H. M. Hodges, Phys. Rev. D 47, 456 (1993); \\
Z. G. Berezhiani, A. D. Dolgov and R. N. Mohapatra, 
Phys. Lett. B 375, 26 (1996) [hep-ph/9511221];\\
V. Berezinsky and A. Vilenkin, Phys. Rev. D 62, 083512 (2000)
[hep-ph/9908257].

\bibitem{baryo-lepto}
L. Bento and Z. Berezhiani, Phys. Rev. Lett. 87, 231304 (2001)
[hep-ph/0107281]; \\
L. Bento and Z. Berezhiani, hep-ph/0111116; \\
L. Bento and Z. Berezhiani, Fortsch. Phys. 50 (2002) 489; \\
R. Foot and R. R. Volkas, Phys. Rev. D 68, 021304 (2003) [hep-ph/0304261]; \\
R. Foot and R. R. Volkas, Phys. Rev. D 69, 123510 (2004) [hep-ph/0402267].

\bibitem{wmap-par} 
D.N. Spergel {\it at al.}, 
Astrophys. J. Suppl. 148, 175 (2003) (astro-ph/0302209).

\bibitem{mabert} 
C. Ma \& E. Bertschinger, Astrophys. J. 455, 7 (1995) [astro-ph/9506072].

\bibitem{bunnwhite} 
E.F. Bunn \& M. White, {\sl Astrophys. J.} {\bf 480}, 6 (1997) 
[astro-ph/9607060].

\bibitem{wmap-data} 
G. Hinshaw {et al.}, Astrophys. J. Suppl. 148 (2003) 135 [astro-ph/0302217].

\bibitem{acbar-data} 
C.L. Kuo {et al.}, Astrophys. J. 600, 32 (2004) [astro-ph/0212289].

\bibitem{2df-teg} 
M. Tegmark, A.J.S. Hamilton, Y. Xu, MNRAS 335 (2002) 887 
[astro-ph/0111575].

\end{thebibliography}
\end{document}